\documentclass[twocolumn,aps,prd,nofootinbib]{revtex4-2}
\pdfoutput=1
\usepackage{amsmath,amsfonts}
\usepackage{epsfig}
\usepackage{caption}
\usepackage{subcaption}
\usepackage{ifpdf}
\usepackage{hyperref}
\usepackage{amsmath}
\usepackage{graphicx}
\usepackage{color}
\usepackage{natbib}
\usepackage{multirow}
\usepackage{kotex}
\usepackage{times}
\usepackage{booktabs,tabularx,dcolumn}

\newcommand{\deff}{d_{\rm eff}}
\newcommand{\msun}{M_\odot}
\newcommand{\be}{\begin{equation}}
\newcommand{\ee}{\end{equation}}
\newcommand{\bea}{\begin{eqnarray}}
\newcommand{\eea}{\end{eqnarray}}

\newcommand{\etal}{{\it et al.}}

\begin{document}

\title{Measurability of neutron star tidal deformability from merging neutron star-black hole binaries}

\author{Hee-Suk Cho}
\email{chohs1439@pusan.ac.kr}
\affiliation{Department of Physics, Pusan National University, Busan, 46241, Korea}
\affiliation{Extreme Physics Institute, Pusan National University Busan, 46241, Korea}

\date{\today}

\begin{abstract}

The neutron star-black hole binary (NSBH) system has been considered one of the promising detection candidates 
for ground-based gravitational-wave (GW) detectors such as LIGO and Virgo.
The tidal effects of neutron stars (NSs) are imprinted on the GW signals emitted from NSBHs as well as binary neutron stars.
The NS tidal deformability ($\lambda_{\rm NS}$) was successfully measured by the binary neutron star signal GW170817
but could not be constrained in the analysis of the two NSBH signals GW200105 and GW200115
due to the low signal-to-noise ratio.
In this work, we study how accurately the parameter $\lambda_{\rm NS}$ can be measured in GW parameter estimation for NSBH signals.
We set the parameter range for the NSBH sources to $[4\msun, 10\msun]$ for the black hole mass,
$[1\msun, 2\msun]$ for the NS mass, and $[-0.9, 0.9]$ for the dimensionless black hole spin.
For realistic populations of sources distributed in different parameter spaces, 
we calculate the measurement errors of $\lambda_{\rm NS}$ ($\sigma_{\lambda_{\rm NS}}$) using the Fisher matrix method.
In particular, we perform a single-detector analysis using the advanced LIGO and the Cosmic Explorer detectors
and a multi-detector analysis using the 2G 
(advanced LIGO-Hanford, advanced LIGO-Livingstone, advanced Virgo, and KAGRA) 
and the 3G (Einstein Telescope and Cosmic Explorer) networks.
We show the distribution of $\sigma_{\lambda_{\rm NS}}$ for the population of sources as a one-dimensional probability density function.
Our result shows that the probability density function curves are similar in shape between advanced LIGO and Cosmic Explorer, 
but Cosmic Explorer can achieve $\sim 15$ times better accuracy overall in the measurement of $\lambda_{\rm NS}$.
In the case of the network detectors, 
the probability density functions are maximum at
$\sigma_{\lambda_{\rm NS}} \sim 130$ and $\sim 4$ for the 2G and the 3G networks, respectively,
and the 3G network can achieve $\sim 10$ times better accuracy overall.
Specifically,
we investigate the distribution of $\sigma_{\lambda_{\rm NS}}$  
for $10^3$ Monte Carlo sources in our parameter range with the NS mass fixed to $m_2=1.4\msun$,
and the result shows that if the sources are located at $d_L \simeq 100{\rm Mpc}$, the parameter estimation results for $\sim 80\%$ of the sources 
can distinguish between the theoretical EOS models at the 1--$\sigma$ level, using the 3G network.
Additionally, we demonstrate that our PDF results are almost unaffected by different choices of the true value of $\lambda_{\rm NS}$.

\end{abstract}

%\pacs{04.30.--w, 04.80.Nn, 95.55.Ym}

\maketitle
%=======	Title		================================	

%=======	Intro		================================	
\section{Introduction}
Since the first gravitational-wave (GW) signal was detected in 2015 \cite{GW150914}, 
the network of the two advanced LIGO (aLIGO) \cite{ALIGO} and advanced Virgo \cite{AVirgo} detectors has observed $90$
GW candidates \cite{GWTC-1,GWTC-2,GWTC-2.1,GWTC-3} through three observing runs.
All GW signals were emitted from a compact binary coalescence (CBC) system such as binary black hole (BBH),  
binary neutron star (BNS), and neutron star-black hole binary (NSBH).
Most GW sources originated from BBHs,
and various BH masses and spins were measured from these signals.
The sources of the two signals GW170817 \cite{GW170817,GW170817PE} and GW190425 \cite{GW190425} 
were identified as BNSs,
and GW170817 enabled us to directly measure the NS tidal deformability
for the first time through observation means.
In particular, since GW170817 had a high signal-to-noise ratio (SNR) $\sim 32$, 
it could be inferred that the soft equation-of-state (EOS) model is preferred over the stiff EOS model \cite{GW170817PE}.
In parameter estimation for BNS signals, a well-constrained tidal parameter is the effective tidal deformability 
rather than the component tidal deformability.
The effective tidal deformability is defined by the combination of the masses ($m_i$) and the component tidal parameters ($\lambda_i$).
Since $\lambda_1$ and $\lambda_2$ are generally strongly correlated,
their measurement errors can be large even though the effective tidal parameter is well constrained as shown in the result of GW170817 \cite{GW170817PE}.

On the other hand, the two NSBH signals GW200105 and GW200115 
were also captured by the LIGO-Virgo network during the third observing run \cite{NSBH_detection}
(for a brief overview of NSBH mergers, refer to \cite{10.3389/fspas.2020.00046}).
Since the GWs from NSBHs also contain an NS tidal effect, information on the tidal parameter can be extracted from those signals.
However, the contribution of tidal deformability to the waveform of the NSBH system is relatively small 
compared to that of the BNS system, especially when the BH mass is much larger than the NS mass.
Therefore, a sufficiently high SNR is required to measure the tidal deformability from the NSBH signals.
Unfortunately, the observed NSBH signals were not able to constrain the tidal deformability of the NSs well due to their low SNRs.
Meanwhile, a large advantage of the NSBH signals when measuring the tidal parameter is that
the individual NS tidal deformability rather than the effective tidal deformability can be obtained directly through parameter estimation
because the tidal deformability of BH is zero.

The purpose of this work is to investigate how accurately the NS tidal deformability ($\lambda_{\rm NS}$) can be measured from NSBH signals.
To this end, we utilize the Fisher matrix method implemented in the python package {\bf GWBench} \cite{Borhanian_2021}
and calculate the measurement errors of $\lambda_{\rm NS}$  for realistic populations of NSBH sources.
Several parameter estimation studies on the measurability of the NS tidal deformability
have been done using Bayesian analysis with
stochastic sampling on NSBH systems as well as Fisher Matrix studies. 
Lackey \etal \cite{PhysRevD.85.044061} estimated tidal deformability from NSBH systems with nonspinning BHs
by applying the Fisher matrix method
to the hybrid waveforms based on BBH waveforms calibrated to NSBH numerical simulations,
and they extended it to the spinning NSBH systems in subsequent work \cite{PhysRevD.89.043009}.
In the former, they showed that for a single aLIGO detector, the tidal parameter $\lambda_{\rm NS}$ can be extracted to $10–40\%$ accuracy from single events
for mass ratios of $q = 2$ and $3$ at a distance of $100 {\rm Mpc}$,
and in the latter, for $q = 2$--$5$,  BH spins $\chi_{\rm BH}=-0.5$--$ 0.75$, NS masses $m_{\rm NS}=1.2\msun $--$ 1.45\msun$, and a distance of $100 {\rm Mpc}$,
 a single aLIGO detector can measure $\lambda_{\rm NS}$ to a $1$--$\sigma$ uncertainty of $\sim 10$--$100\%$.
 For both works, they showed that the uncertainty in $\lambda_{\rm NS}$ is an order of magnitude smaller for the 3G detector Einstein Telescope.
 On the other hand, Kumar \etal \cite{PhysRevD.95.044039} performed a Bayesian analysis of NSBH systems with spinning BHs to study the measurability of $\lambda_{\rm NS}$
with aLIGO and found that 20--35 events can constrain $\lambda_{\rm NS}$
 within $25$--$50\%$, depending on the EOS.

This paper is organized as follows.
In Sec. \ref{sec.method}, we introduce various waveform models
recently developed for use in GW data analysis for BNS or NSBH systems
based on the Effective-One-Body (EOB) formalism and the phenomenological fit approach.
Next, we give a brief overview of the Bayesian parameter estimation and the Fisher matrix approach
in terms of parameter measurement accuracy
and list the 2G and the 3G detectors used in our analysis.
The results are given in Sec. \ref{sec.result}.
First, we compare the parameter measurement errors obtained from the Fisher matrix approach 
with those obtained from the Bayesian parameter estimation simulations and verify the reliability of the Fisher matrix method in our analysis.
Next, we investigate the suitability of four recent waveform models for applying the Fisher matrix method
to NSBH sources in our parameter range
and adopt a representative waveform model (denoted by SEOBNR\_T in this work).
Then, we apply the Fisher matrix method to SEOBNR\_T
and calculate the measurement errors of $\lambda_{\rm NS}$ for our NSBH sources.
We perform a single-detector analysis as well as a multi-detector analysis
and provide comparisons between the 2G and the 3G detectors.
In particular, using the results for the 3G network, we present a specific example describing
how well the theoretical EOS models can be constrained by the NSBH signals.
In Sec. \ref{sec.summary}, we summarize our results and provide some discussion.

\section{Method} \label{sec.method}

\subsection{Waveform models}

To simulate an NSBH signal, the waveform model requires the full Inspiral-Merger-Ringdown (IMR) expression, including the NS tidal effect.
To date, various IMR waveform models have been developed and implemented in LAL (LIGO Algorithm Library \cite{lalsuite}) for use in GW data analysis.
Those models are roughly classified into two waveform families according to the construction formalism.
One is based on the Effective-One-Body (EOB) formalism and the other is based on the phenomenological fit approach.
In both approaches, the waveform from the late inspiral to the merger-ringdown is calibrated against the aligned-spin BBH NR waveforms.
So they are represented by the ``SEOBNR" and ``IMRPhenom" models.

The EOB formalism is basically constructed in the time domain.
For computational efficiency, the frequency-domain model SEOBNRv4\_ROM \cite{P_rrer_2014,PhysRevD.95.044028} 
has been developed based on the time-domain model SEOBNRv4 using a reduced-order-quadrature rule \cite{PhysRevD.94.044031}.
SEOBNRv4\_ROM\_NRTidalv2 builds on SEOBNRv4\_ROM by adding tidal correction terms that are constructed from high-resolution BNS NR simulations \cite{PhysRevD.96.121501,PhysRevD.100.044003}.
SEOBNRv4\_ROM\_NRTidalv2\_NSBH \cite{PhysRevD.102.043023} was built from SEOBNRv4\_ROM\_NRTidalv2 to generate aligned-spin NSBH waveforms 
by adding corrections to the wave amplitude \cite{PhysRevD.92.084050}.

Meanwhile, the IMRPhenom models are defined in the frequency domain.
Early versions of the IMRPhenom models were developed for the BBH system.
IMRPhenomPv2 has been mainly used for CBC analyses in recent years.
This model is based on the precessing-spin model IMRPhenomP \cite{PhysRevLett.113.151101}
and the aligned-spin model IMRPhenomD \cite{PhysRevD.93.044006,PhysRevD.93.044007}.
IMRPhenomPv2\_NRTidalv2 is based on IMRPhenomPv2
and includes the same tidal correction terms \cite{PhysRevD.96.121501,PhysRevD.100.044003} as in SEOBNRv4\_ROM\_NRTidalv2.
The IMRPhenom family also has an NSBH model IMRPhenomNSBH \cite{PhysRevD.101.124059}.
This model is based on the amplitude of IMRPhenomC \cite{PhysRevD.82.064016} and the phase of IMRPhenomD \cite{PhysRevD.93.044006,PhysRevD.93.044007},
and incorporates NS tidal effects \cite{PhysRevD.100.044003} and amplitude corrections \cite{PhysRevD.92.084050} similar to SEOBNRv4\_ROM\_NRTidalv2\_NSBH.

In this work, four recent IMR waveform models containing NS tidal effects are considered,
which are listed in Table \ref{tab.models}.
Finally, we note that the TaylorF2 model also includes NS tidal effects but generates inspiral-only waveforms.
TaylorF2 is reliable in parameter estimation for BNS systems such as GW170817 \cite{GW170817PE} and GW100425 \cite{GW190425}.
However, after some consistency tests, we have verified that TaylorF2 is not suitable for the analysis of our NSBH sources.

\begin{table}[t]
\centering
\begin{tabular}{c cc}
\hline\hline
Full name (implemented in LAL) & \multicolumn{2}{c}{References}   \\
Short label (used in this work)&    Base model & Corrections   \\
\hline
SEOBNRv4\_ROM\_NRTidalv2  &   \cite{PhysRevD.94.044031, P_rrer_2014,PhysRevD.95.044028} & \cite{PhysRevD.96.121501,PhysRevD.100.044003}   \\
SEOBNR\_T &&\\
SEOBNRv4\_ROM\_NRTidalv2\_NSBH \cite{PhysRevD.102.043023}& \cite{PhysRevD.94.044031, P_rrer_2014,PhysRevD.95.044028} & \cite{PhysRevD.96.121501,PhysRevD.100.044003,PhysRevD.92.084050} \\
 SEOBNR\_NSBH &&\\
IMRPhenomPv2\_NRTidalv2  & \cite{PhysRevLett.113.151101,PhysRevD.93.044006,PhysRevD.93.044007} & \cite{PhysRevD.96.121501,PhysRevD.100.044003} \\
IMRPhenomP\_T &&\\
IMRPhenomNSBH \cite{PhysRevD.101.124059}   &\cite{PhysRevD.82.064016,PhysRevD.93.044006,PhysRevD.93.044007} & \cite{PhysRevD.100.044003,PhysRevD.92.084050} \\
IMRPhenom\_NSBH &&\\
\hline\hline
\end{tabular}
\caption{Waveform models used in our analysis for NSBH systems.}
\label{tab.models}
\end{table}

\subsection{Bayesian parameter estimation}

The physical properties of the GW source can be measured in the parameter estimation procedure \cite{PhysRevD.91.042003}.
This process is based on Bayesian inference statistics,
and the algorithm explores the entire parameter space,
computing the overlaps between model waveforms and detector data.
The result of Bayesian parameter estimation can be given as the posterior probability density functions (PDFs) of the parameters considered.
Given the detector data $x$ containing the GW signal $s$ and noise $n$,
the overlap between $x$ and the model waveform $h$ is defined as
\be \label{eq.overlap}
\langle x | h \rangle = 4 {\rm Re} \int_0^{\infty}  \frac{\tilde{x}(f)\tilde{h}^*(f)}{S_n(f)} df,
\ee
where the tilde denotes the Fourier transform of the time-domain
waveform, $S_n(f)$ is the detector's noise power spectral density (PSD).
For efficiency, the integration is performed in the frequency range $[f_{\rm min}, f_{\rm max}]$,
and the choice of these frequencies depends on the PSD curve.

The Bayesian posterior probability that the GW signal $s$ contained in the data $x$ is characterized by the parameters $\theta$, 
where $\theta$ is the set of parameters considered in the analysis,
can be given by the prior $p(\theta)$ and the likelihood $L(x|\theta)$ as
\be \label{eq.posterior}
p(\theta|x) \propto p(\theta)L(x|\theta).
\ee
The likelihood is given as \cite{PhysRevD.46.5236,PhysRevD.49.2658}
\bea \label{eq.likelihood}
L(x|\theta) &\propto& {\rm exp}\bigg[-\frac{1}2 \langle x-h(\theta)|x-h(\theta)\rangle\bigg] \\
                        &=& {\rm exp}\bigg[-\frac{1}2 \langle s+n-h(\theta)|s+n-h(\theta)\rangle\bigg]. \\
\eea
If the signal is strong enough (i.e., high SNR limit), the noise can be removed from the above equation,
giving the likelihood function as
\be \label{eq.likelihood-2}
L(\theta)\propto {\rm exp} \bigg[ -\frac{1}2\{\langle s|s \rangle+\langle h(\theta)|h(\theta)\rangle-2\langle s|h(\theta)\rangle\} \bigg].
\ee
Employing the definition of SNR $\rho=\sqrt{\langle s|s \rangle}$ \cite{PhysRevD.85.122006},
the above equation can be re-written as
\be \label{eq.likelihood-3}
L(\theta) \propto {\rm exp} [-\rho^2\{1- \langle \hat{s}|\hat{h}(\theta)\rangle\}],
\ee
where $\hat{h}\equiv h/\rho$,
and we assume that the waveform model can describe the signal waveform almost exactly, 
i.e., $h(\theta_0) \simeq s$ (where $\theta_0$ represents the true parameter values), 
hence $\langle h(\theta_0)|h(\theta_0) \rangle \simeq  \langle s|s \rangle= \rho^2$.
In the above equation, the term $ \langle \hat{s}|\hat{h}(\theta)\rangle$ represents the distribution of the normalized overlaps
between the signal and the model waveforms, and this overlap distribution is maximum at $\theta=\theta_0$.
Therefore, the shape of the likelihood surface can be given by the overlap distribution, 
and its scale of interest depends on the SNR \cite{PhysRevD.87.024004}. 
At small scales (i.e., high SNRs), the shape of the overlap distribution is nearly quadratic around the maximum position.
If we assume a flat prior in Eq. \ref{eq.posterior}, the posterior distribution is equivalent to the likelihood distribution.
Therefore, in the high SNR limit, the posterior PDF follows a multivariate Gaussian distribution centered around the position of the true parameter values.

\subsection{Fisher matrix}

 If a multivariate Gaussian function is represented by
 \be
 f(x) = \exp(-\Sigma^{-1}_{ij}  x_i x_j/2),
 \ee
$\Sigma_{ij}$ corresponds to the covariance matrix, and its inverse matrix represents the Fisher matrix ($\Gamma_{ij}$).
Therefore, given the Gaussian function above,
the Fisher matrix can be obtained by
\be
\Gamma_{ij}=-\frac{\partial^2 \ln f(x)}{\partial x_i  \partial x_j}.
\ee
Analogously, since the likelihood in Eq. \ref{eq.likelihood-3} follows a multivariate Gaussian distribution,
the corresponding Fisher matrix can be given by
\be      \label{eq.covariance matrix}
\Gamma_{ij} = - \frac{\partial^2 {\rm ln} L(\theta)}{\partial \theta_i \partial \theta_j} \bigg|_{\theta=\theta_0} 
=-\rho^2 {\partial^2   \langle \hat{s}|\hat{h}(\theta)\rangle \over \partial \theta_i \partial \theta_j}\bigg|_{\theta=\theta_0}.
\ee
Thus, the Fisher matrix describes the curvature of the log-likelihood or the overlap surface at the position of the true parameter values,
Furthermore, the second equality means that a specific iso-match contour in the overlap surface
corresponds to a specific confidence region of the likelihood distribution for a given SNR \cite{PhysRevD.87.024035,Cho_2014,Cho_2015}.
Using the relation $\hat{h}\equiv h/\rho$, the above formula can equivalently be written as \cite{PhysRevD.49.1723,PhysRevD.77.042001,Porter_2002}
\be\label{eq.FM}
\Gamma_{ij} = -\rho^2 {\partial^2   \langle \hat{s}|\hat{h}(\theta)\rangle \over \partial \theta_i \partial \theta_j}\bigg|_{\theta=\theta_0}
=\bigg \langle {\partial h(\theta) \over \partial \theta_i} \bigg| {\partial h(\theta) \over \partial \theta_j} \bigg \rangle\bigg|_{\theta=\theta_0}.
\ee
The last term is the familiar expression of the Fisher matrix.
In a multivariate Gaussian distribution, the measurement error ($\sigma_i$) 
and the correlation coefficient ($C_{ij}$) can be obtained from the Fisher matrix as
\be \label{eq.sigma}
\sigma_i=\sqrt{(\Gamma^{-1})_{ii}},  \ \ \ C_{ij}={(\Gamma^{-1})_{ij} \over \sqrt{(\Gamma^{-1})_{ii} (\Gamma^{-1})_{jj}}}.
\ee

Since the Fisher matrix has a simple functional form and is easy to apply to analytical waveform models,  
this approach has been mainly used in many past works since it was introduced in \cite{PhysRevD.49.2658, PhysRevD.52.848}.
However, the Fisher matrix has some well-known limitations (for detailed reviews, refer to \cite{PhysRevD.77.042001}).
First of all, the Fisher matrix is only reliable at high SNRs
because it is derived with a high SNR assumption as described above.
In addition, since the computation of the Fisher matrix is entirely dependent on the waveform model $h(\theta)$ as in Eq. \ref{eq.FM},
the results highly rely on the accuracy of the model used.
Some issues induced by using the inspiral-only waveform model TaylorF2 have been 
thoroughly studied in past works \cite{PhysRevD.88.084013,Mandel_2014,Cho_2014}.
Recently, Harry and Lundgren \cite{PhysRevD.104.043008} pointed out that 
the TaylorF2-applied Fisher matrix is unsuitable to predict the match between two BNS waveforms
when including tidal terms.
Another well-known limitation of the Fisher matrix is the poor applicability of prior information.
Bayesian parameter estimation allows for all forms of prior information, 
while only Gaussian prior functions can be applied analytically to the Fisher matrix method  \cite{PhysRevD.49.2658, PhysRevD.52.848}.
Cho \cite{Cho_2022} showed that the measurement error of the intrinsic parameters can be reduced to $\sim 70\%$
of the original priorless error ($\sigma^{\rm priorless}_{\theta}$) if the standard deviation of the Gaussian prior is
similar to $\sigma^{\rm priorless}_{\theta}$,
and thus the prior effect can be ignored at sufficiently high SNRs.
Therefore, we choose the IMR waveform model and the high SNR
in our analysis to avoid the above limitations.

To describe the waveforms of an aligned-spin NSBH system,
five extrinsic parameters (true distance $d_L$, orbital inclination $\theta_{JN}$, 
polarization angle $\Psi$, and sky position angles RA, DEC),
five intrinsic parameters (two masses $m_1, m_2$, dimensionless spins $\chi_{\rm BH}, \chi_{\rm NS}$, and dimensionless NS tidal deformability $\lambda_{\rm NS}$),
and two arbitrary constants (coalescence time $t_c$ and coalescence phase $\phi_c$) are required.
Since the NS mass is much smaller than the BH mass 
and the NS spins observed from BNS systems are very small ($\chi_{\rm NS} \lesssim 0.05$) \cite{Burgay:2003ur, Stovall_2018},
the NS spin has a negligible contribution to the wave phase and thus has no effect on our analysis.
Therefore, for simplicity, we assume the NS spin to be zero and consider only the BH spin ($\chi_{\rm BH}$) in this work.
In addition, we adopt the soft EOS model APR4 \cite{PhysRevD.79.124032},
which is one of the most preferred models in the parameter estimation results for GW170817 \cite{GW170817PE},
to choose the true value of $\lambda_{\rm NS}$.

A GW waveform can be described as
\be \label{eq.hf}
h(f)=A(f) e^{i \psi(f)}.
\ee
The signal strength (i.e., SNR) is entirely governed by the wave amplitude ($A$),
and the amplitude is given by the extrinsic parameters 
and the chirp mass ($M_c \equiv (m_1+m_2)\eta^{3/5}$,
where $\eta \equiv (m_1 m_2)/(m_1+m_2)^2$ is the symmetric mass ratio).
The wave phase $\psi(f)$ is only a function of the intrinsic parameters and $t_c$ and  $\phi_c$.
The true values of $t_c$ and $\phi_c$ can be arbitrarily selected, and their choice 
does not affect the measurement accuracy of other parameters.
However, since these two parameters are strongly correlated with the intrinsic parameters,
they must be considered variables when constructing the Fisher matrix (e.g., see Table A1 of \cite{Cho_2022}).
On the other hand, the extrinsic parameters are strongly correlated with each other
but weakly correlated with the intrinsic parameters.
Thus, when focusing on the intrinsic parameters,
it is very efficient to use a single effective parameter that represents the five extrinsic parameters,
and we use the parameter $\deff$ (effective distance \cite{PhysRevD.85.122006}) in this work.
At leading order, the wave amplitude can be given by $A \propto M_c^{5/6}/ \deff$.
For fixed $\deff$, the measurement errors of the intrinsic parameters are independent of 
the choice of the individual extrinsic parameters, so the extrinsic parameters are not considered
variables in our Fisher matrix.
Therefore, in this work, the Fisher matrix can be given by a $6 \times 6$ matrix with the components $\{M_c,\eta,\chi_{\rm BH},\lambda_{\rm NS}, t_c,\phi_c\}$.

\subsection{Detectors}

\begin{figure}[t]
\begin{center}
\includegraphics[width=\columnwidth]{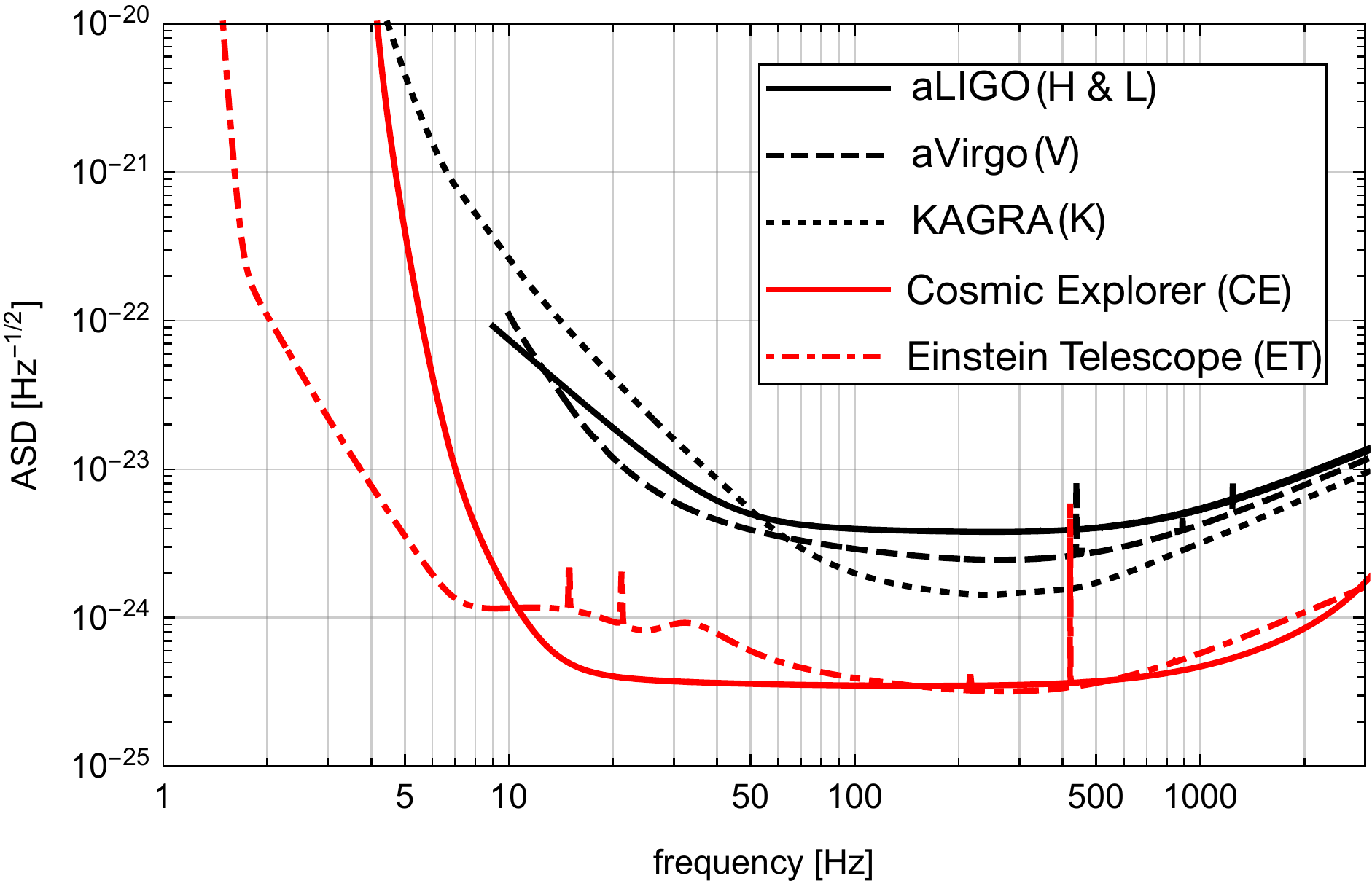}
\caption{\label{fig.PSD} Amplitude spectral densities (ASDs) $\sqrt{S_n(f)}$ of the 2G and 3G detectors used in this work. 
The frequency range is set to $f_{\rm min}= 10  (5)$ Hz for the 2(3)G detectors
and $f_{\rm max}= 2048$ Hz for all detectors.}
\end{center}
\end{figure}

We consider the four 2G GW detectors, aLIGO-Hanford (H) and Livingstone (L) \cite{ALIGO}, advanced Virgo (V) \cite{AVirgo}, and KAGRA (K) \cite{PhysRevD.88.043007},
and the two 3G detectors, Cosmic Explorer (CE) \cite{Abbott_2017} and Einstein Telescope (ET) \cite{Punturo_2010}.
The sensitivity curves of the detectors are shown in Fig. \ref{fig.PSD}.
These PSDs are available in {\bf GWBench} \cite{Borhanian_2021},
labeled aLIGO (H \& L), V+ (V), K+ (K), ET (ET), and CE1-40-CBO (CE).
We assume $f_{\rm min}= 10$ and 5 Hz for the 2G and the 3G detectors, respectively,
and $f_{\rm max}= 2048$ Hz for all detectors.
The locations (longitude and latitude) and the orientations (orientation of the y-arm with respect to due East) 
of the detectors are summarized in Table III and Fig. 4 of \cite{Borhanian_2021}.
Note that H and L have the same PSD curve but their locations and orientations are different.
CE (Idaho, USA) is located at a site similar to H (Washington, USA) but has a different orientation.
ET is set to the same coordinates as V (Cascina, Italy) and consists of three V-shaped detectors, ET1, ET2, and ET3,
that form an equilateral triangle, and one of them has the same orientation as V.

\section{Result}  \label{sec.result}

\subsection{Comparison between Bayesian parameter estimation and Fisher matrix}

\begin{figure*}[t]
\begin{center}
\includegraphics[width=2\columnwidth]{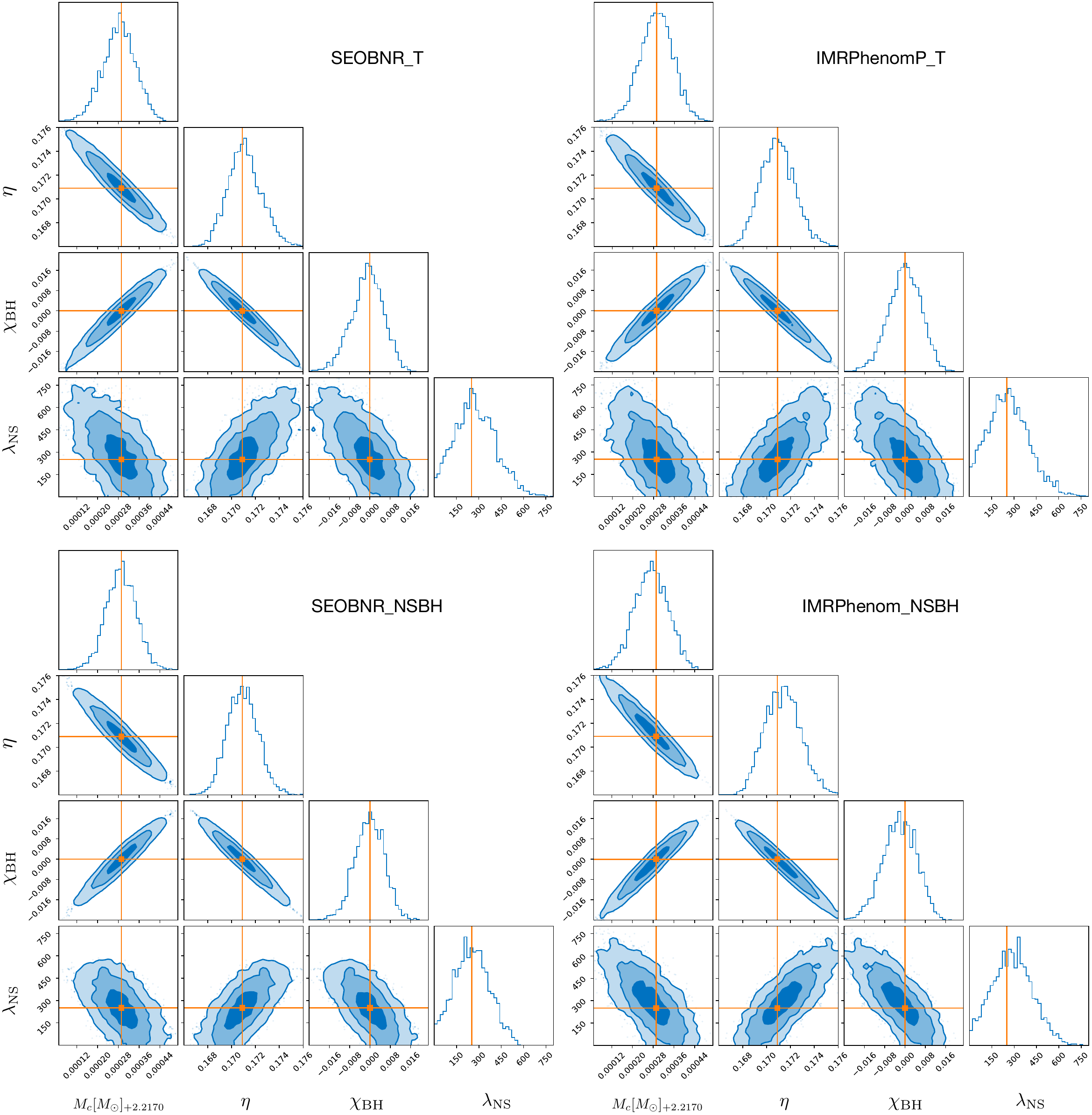}
\caption{\label{fig.posteriors} Posteriors for our fiducial NSBH source with the true values $\{m_1,m_2,\chi_{\rm BH},\lambda_{\rm NS}, t_c, \phi_c  \} = \{5\msun ,1.4\msun,0,251,0,0\}$.
We use the aLIGO PSD and assume $\rho=200$. The 2--D contours correspond to 39, 86, and $99\%$ confidence regions. Note that the results for $t_c$ and $\phi_c$ are omitted here.}
\end{center}
\end{figure*}

\begin{figure}[t]
\begin{center}
\includegraphics[width=\columnwidth]{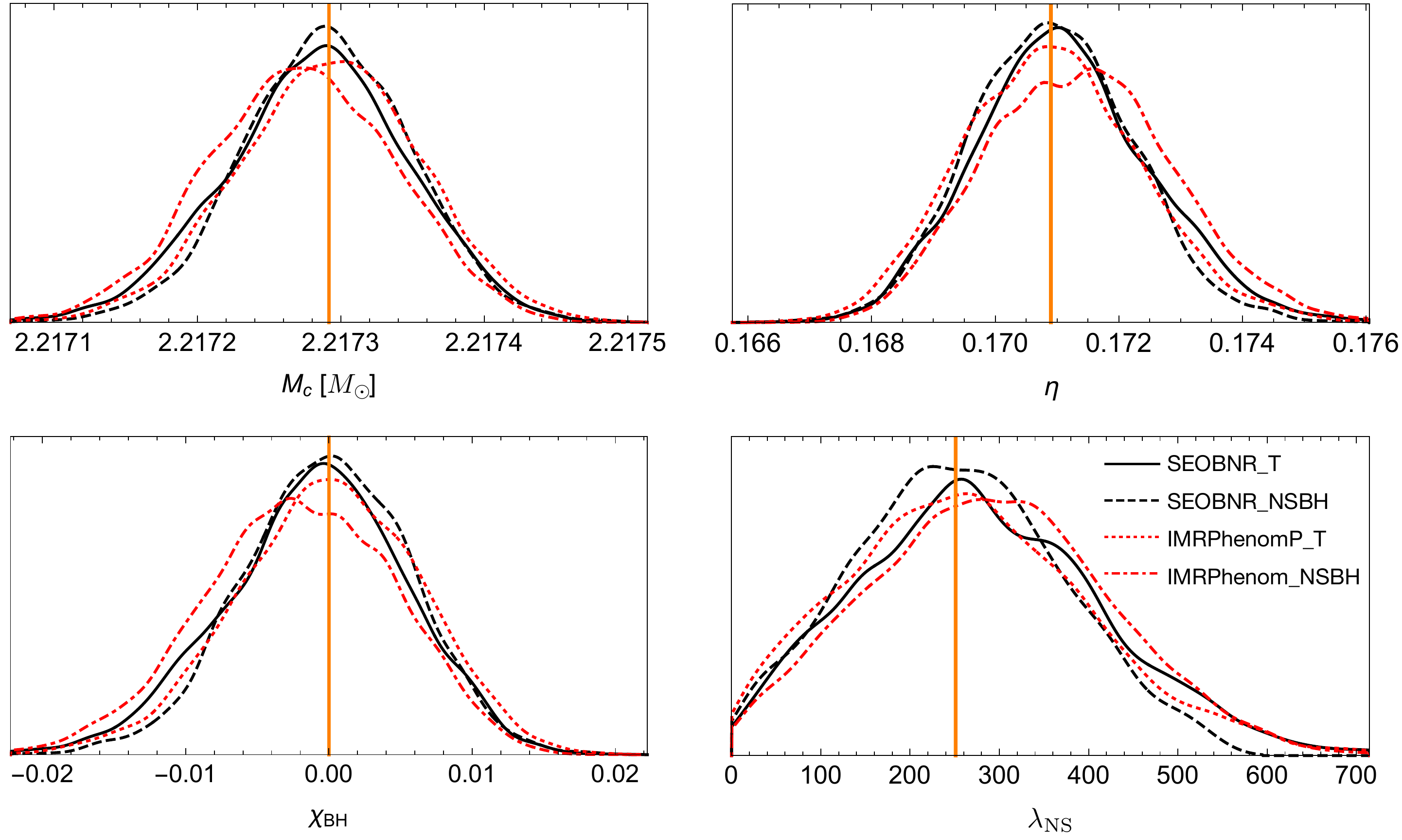}
\caption{\label{fig.PDF-comparison} Comparison of the PDFs between the waveform models.
All PDF curves show similar confidence intervals.}
\end{center}
\end{figure}

We assume our fiducial NSBH source with the true values $\{m_1,m_2,\chi_{\rm BH},\lambda_{\rm NS}, t_c, \phi_c  \} = \{5\msun ,1.4\msun,0,251,0,0\}$.
Here, the true value of $\lambda_{\rm NS}$ is determined by the NS mass ($m_2$) according to the APR4 EOS model.
We inject the fiducial NSBH signal into the aLIGO PSD\footnote{For our fiducial binary system, the time to merger from $f_{\rm min}=5{\rm Hz}$ is about 2400 seconds, which is about 6.5 times longer than that from $f_{\rm min}=10{\rm Hz}$,
so a much longer time is required to run parameter estimation for 3G detectors. 
Moreover, the reliability of the Fisher matrix method is almost independent of PSD.
Therefore, in this work, for comparison with the Fisher matrix results, we perform Bayesian parameter estimation only for the aLIGO detector.} 
and perform Bayesian parameter estimation
using the {\bf Bilby} library \cite{Ashton_2019}, 
which is one of the parameter estimation packages.
We use the {\bf Dynesty} nested sampling algorithm  \cite{dynesty:Speagle_2020} 
and the multi-banded likelihood technique described in \cite{PhysRevD.104.044062}.
The parameter estimation algorithm typically explores the entire extrinsic and intrinsic parameter space.
However, since we focus on the intrinsic parameters, 
we fix the extrinsic parameters with their injection values that satisfy $\rho \sim 200$,
thus the algorithm runs in the 6--D space with the parameters $\{M_c,\eta,\chi_{\rm BH},\lambda_{\rm NS}, t_c,\phi_c\}$.
Note that,
even if the extrinsic parameters are considered variables in the parameter estimation runs,
the results of the intrinsic parameters are nearly unchanged,  (e.g., see, Fig. 16 of \cite{PhysRevD.105.124022}).
We assume the flat priors in the ranges 
$[m_i-1\msun, m_i+1\msun]$ for $m_1$ and $m_2$, $[-0.9, 0.9]$ for $\chi_{\rm BH}$, and $[0,  5000]$ for $\lambda_{\rm NS}$.
The priors of $t_c$ and $\phi_c$ are given as $[-1s,+1s]$ and [$-\pi ,+\pi$], respectively.

To verify consistency between the four waveform models,
we perform four parameter estimation runs using those models for the same fiducial NSBH source.
The results for the main intrinsic parameters are displayed in Fig. \ref{fig.posteriors}, 
showing similar posterior distributions for all waveform models.
In each panel, the 2--D contours correspond to 39, 86, and $99\%$ confidence regions, respectively.
The three parameters $M_c, \eta,$ and $\chi_{\rm BH}$ are strongly correlated with each other but weakly correlated with the NS tidal parameter $\lambda_{\rm NS}$.
For direct comparison, we show the 1--D PDF curves for all waveform models together in Fig. \ref{fig.PDF-comparison}.
In each plot, all curves are similar to Gaussian distributions
and show similar confidence intervals (i.e., measurement errors).

\begin{table}[b]
\centering
\begin{tabular}{c cccr}
\hline\hline
Waveform model&$ \sigma_{M_c} [M_{\odot}]$ & $\sigma_{\eta}$ & $\sigma_{\chi_{\rm BH}}$  & $\sigma_{\lambda_{\rm NS}}$   \\
\hline
SEOBNR\_T & 7.16E-5  & 1.61E-3  & 7.21E-3  &  155 \\
SEOBNR\_NSBH & 7.29E-5  & 1.67E-3  & 7.42E-3  &  152\\
IMRPhenomP\_T  & 6.91E-5  & 1.56E-3  & 6.94E-3  &  141\\
IMRPhenom\_NSBH  & 4.86E-5  & 0.96E-3  & 4.44E-3  &  26\\
 \hline\hline
\end{tabular}
\caption{Measurement errors ($\sigma_i$) calculated by the 6--D Fisher matrix for our fiducial NSBH source with $\{m_1,m_2,\chi_{\rm BH},\lambda_{\rm NS}, t_c, \phi_c  \} = \{5\msun ,1.4\msun,0,251,0,0\}$ assuming $\rho=200$.}
\label{tab.FM-error}
\end{table}

\begin{figure}[t]
\begin{center}
\includegraphics[width=\columnwidth]{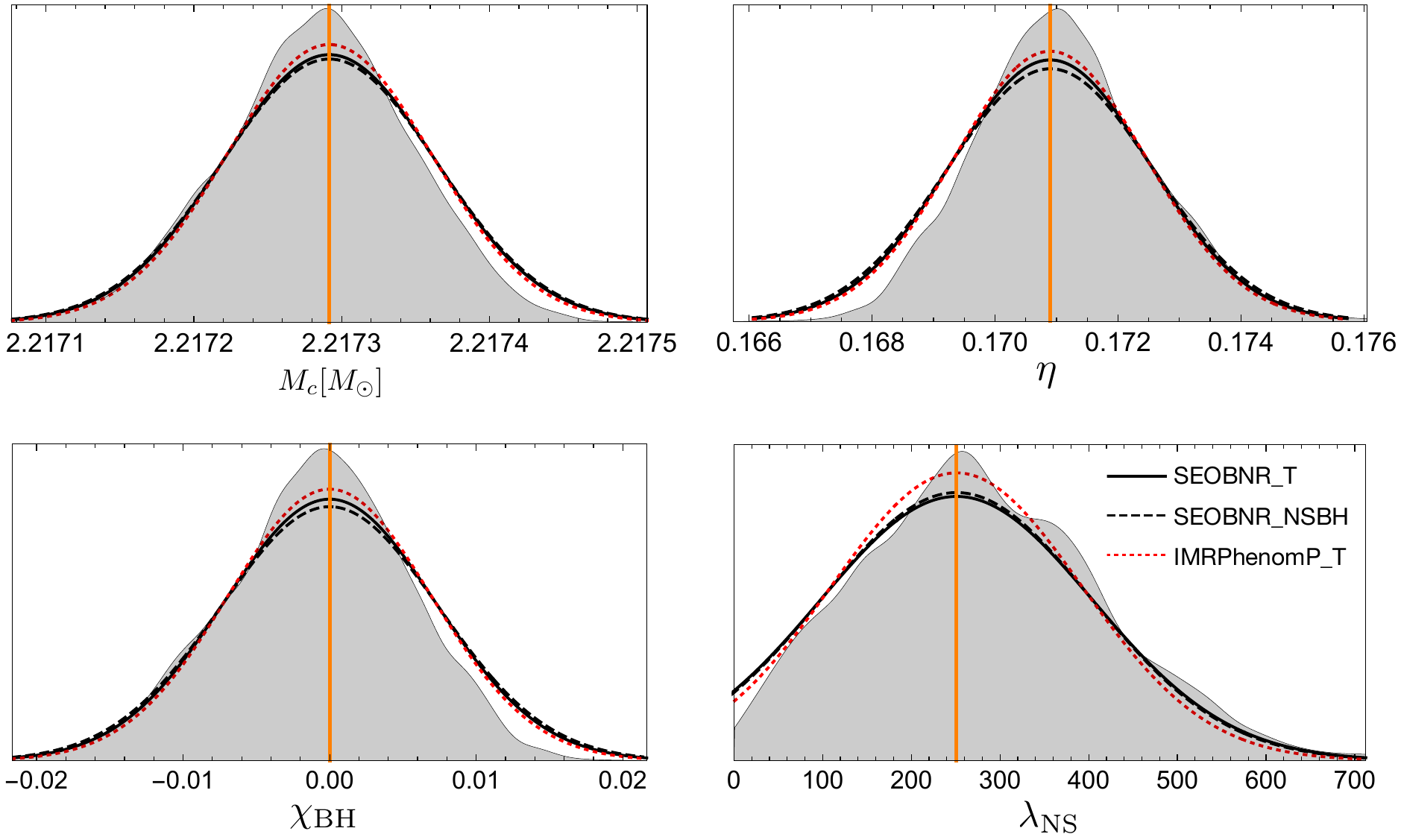}
\caption{\label{fig.FM-Bayesian-comparison} Gaussian PDFs described by the measurement errors in Table \ref{tab.FM-error}. For comparison, we also present the Bayesian posterior PDF for the SEOBNR\_T model (shaded curve).}
\end{center}
\end{figure}

We also calculate the measurement errors for the main intrinsic parameters using the Fisher matrix method.
The results for the four waveform models are listed in Table \ref{tab.FM-error}.
The SEOBNR\_T, SEOBNR\_NSBH, and IMRPhenomP\_T models show very consistent errors for all parameters,
but the IMRPhenom\_NSBH model gives significantly smaller errors compared to the other models, especially for the tidal parameter.
For a measurement error given by the Fisher matrix, one can set a Gaussian PDF with the standard deviation equal to the error.
Figure \ref{fig.FM-Bayesian-comparison} shows the Gaussian PDFs determined by the measurement errors in Table \ref{tab.FM-error},
ignoring the result for IMRPhenom\_NSBH.
For all parameters, the Gaussian curves for the three models are nearly identical.
We also present the Bayesian posterior PDF for the SEOBNR\_T model together with the Gaussian curves.
It can be seen that all posterior PDFs are slightly asymmetric but their maximum positions are unbiased from the true values.
The comparison between the Bayesian and the Gaussian PDF curves shows that
the Fisher matrix and Bayesian parameter estimation give similar results at
high SNRs, which is indeed an underlying assumption for doing any
Fisher matrix study in the first place.

\subsection{Choice of the waveform model: SEOBNR\_T}

\begin{figure*}[t]
\begin{center}
\includegraphics[width=2\columnwidth]{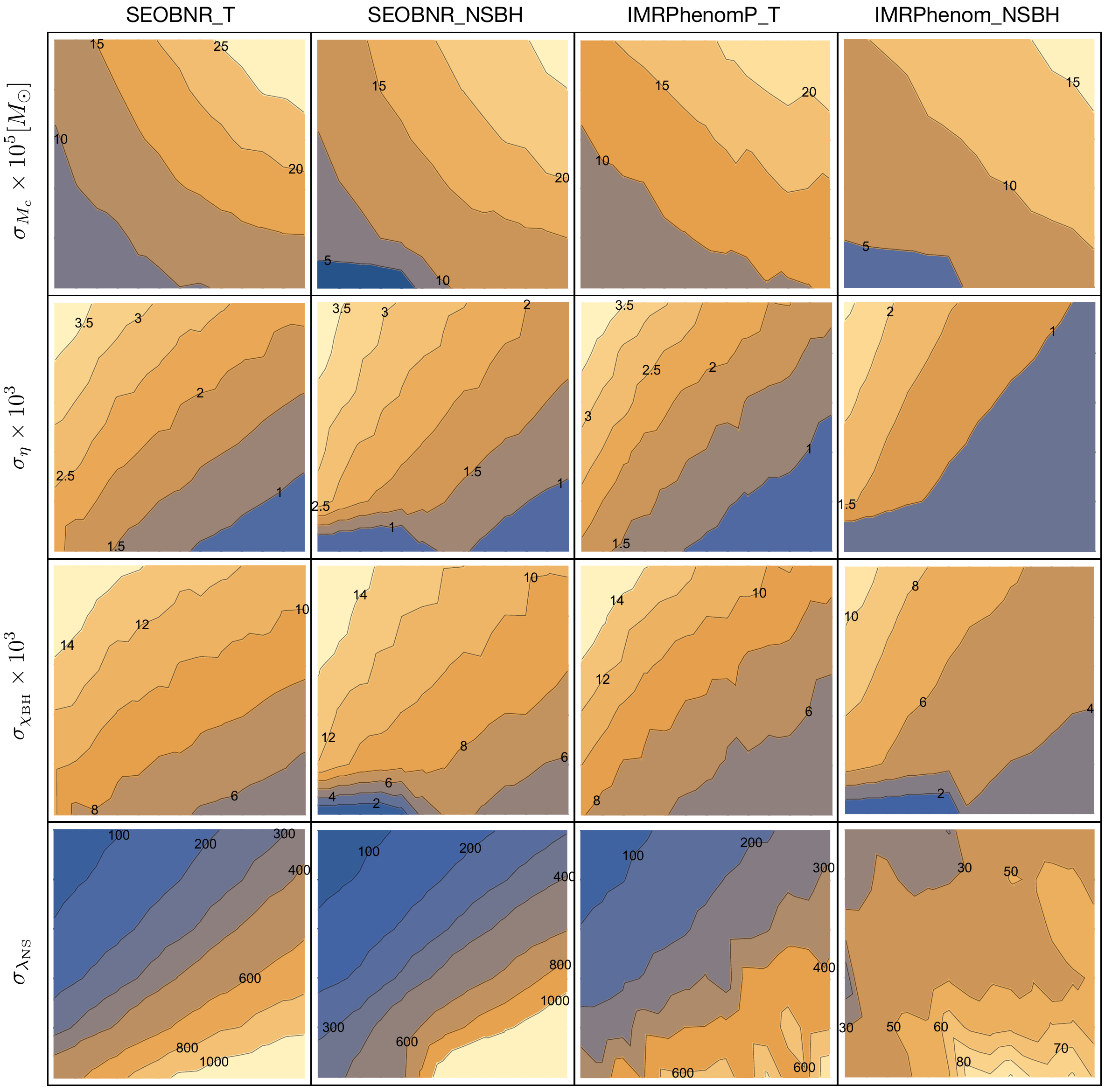}
\caption{\label{fig.model-comparison-4-params} Measurement errors ($\sigma_i$) calculated by the Fisher matrix method using the four waveform models. We use the aLIGO PSD and assume $\chi_{\rm BH}=0$ and $\deff=40{\rm Mpc}$ for all sources.
Each plot is given in the $m_1$--$m_2$ plane in the range $[4\msun, 10\msun]$ for $m_1$ and $[1\msun, 2\msun]$ for $m_2$. }
\end{center}
\end{figure*}

We set our parameter range in the $m_1$--$m_2$ plane
to $[1\msun, 2\msun]$ for the NS mass and $[4\msun, 10 \msun]$ for the BH mass.
For NSBH sources distributed in our parameter space,
we calculate the measurement errors of the intrinsic parameters, $M_c,\eta,\chi_{\rm BH},$ and $\lambda_{\rm NS}$, 
using the Fisher matrix method.
We adopt the aLIGO PSD and assume $\chi_{\rm BH}=0$ and $\deff=40{\rm Mpc}$ \footnote{The condition $\deff=40{\rm Mpc}$ can be simply obtained by choosing the distance $d_L=40{\rm Mpc}$, the optimal sky position (i.e., the direction perpendicular to the plane given by the detector arms), and the optimal orientation ($\Psi=\theta_{JN}=0$).}
for all sources.
Figure \ref{fig.model-comparison-4-params} shows the measurement errors in the $m_1$--$m_2$ plane for the four waveform models.
The first three models show similar trends of contours across the parameter space 
for all parameters, but the IMRPhenom\_NSBH model gives significantly inconsistent results.
Thus, we rule out the IMRPhenom\_NSBH model from our analysis. 
Meanwhile, the results of SEOBNR\_NSBH and IMRPhenom\_NSBH exhibit irregular behavior in the bottom-left corner,
where the measurement errors of the masses and the spin parameters abruptly drop off.
Thus, we also rule out the SEOBNR\_NSBH model from our analysis.
Finally, the SEOBNR\_T and the IMRPhenomP\_T models show relatively consistent error contours,
but the result of SEOBNR\_T looks more clear and has less variation in contours, especially for the tidal parameter.
Therefore, we select SEOBNR\_T as our reference waveform model for our main analysis of the NS tidal deformability.
Although the tidal correction terms of SEOBNR\_T are calibrated to equal-mass BNS
systems  \cite{PhysRevD.96.121501,PhysRevD.100.044003}, the error contours for this model are similar to those for SEOBNR\_NSBH overall
and do not show any irregular behavior in our entire mass range.
Therefore, we conclude that SEOBNR\_T is suitable for our Fisher matrix study.

\begin{figure}[t]
\begin{center}
\includegraphics[width=\columnwidth]{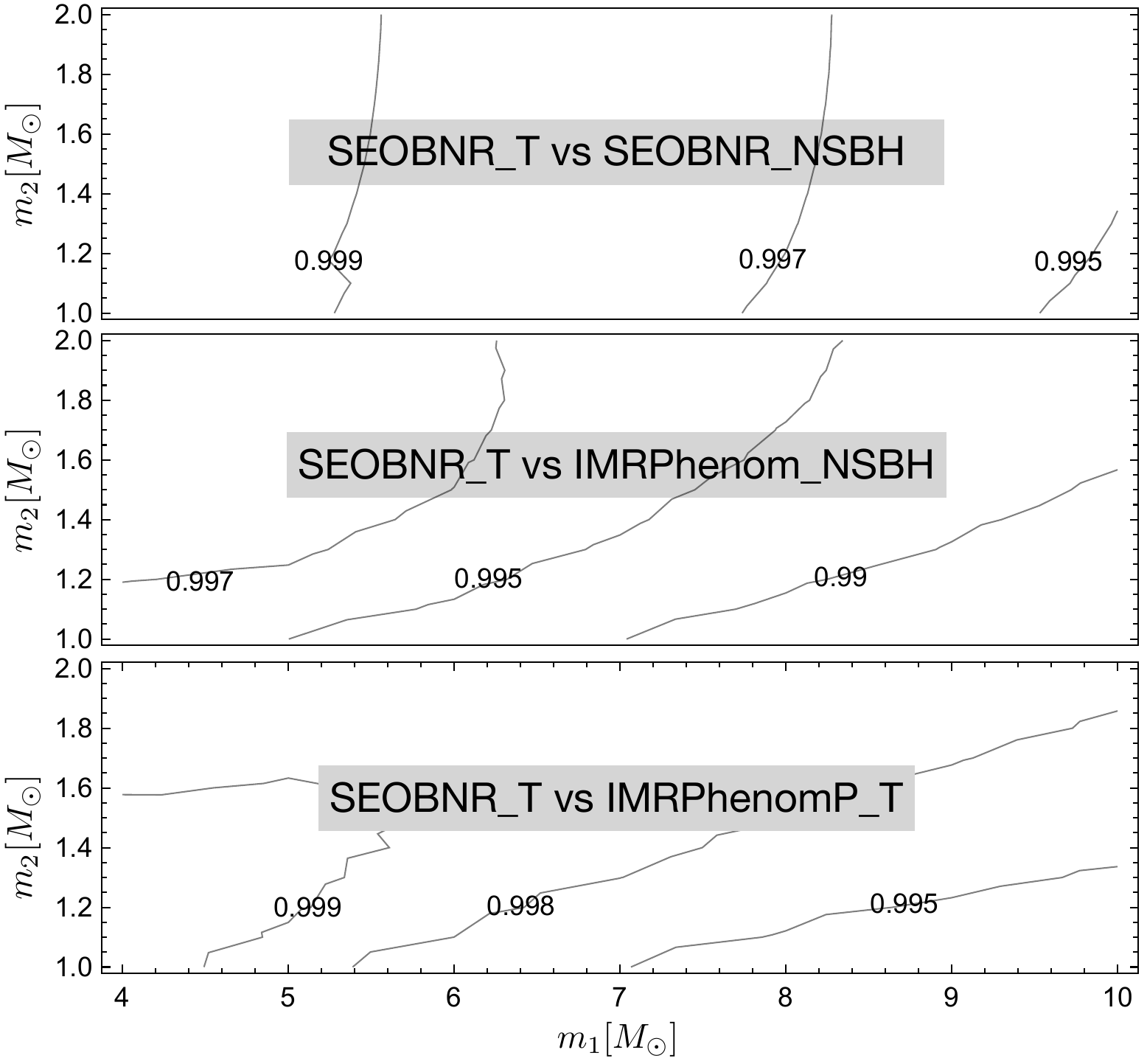}
\caption{\label{fig.overlap} Faithfulness between SEOBNR\_T and the other three models.
We use the aLIGO PSD and assume $\chi_{\rm BH}=0$.
These results do not show any irregular behavior in our entire parameter space.}
\end{center}
\end{figure}

To investigate the origin of the irregular behavior in the two NSBH-dedicated models,
we compare the NSBH-dedicated models with SEOBNR\_T.
To carry out their comparisons, we calculate the {\it faithfulness}, which is determined by
 maximizing the normalized overlap over the coalescence time and phase:
 \be
 {\cal{F}}={\rm max} \frac {\langle h_1(t_c,\phi_c) | h_2 \rangle}{\sqrt{\langle h_1| h_1 \rangle \langle h_2| h_2 \rangle}},
 \ee
where $h_1$ and $h_2$ represent the waveforms of two different models with the same intrinsic parameters. 
The result is given in Fig. \ref{fig.overlap}.
We obtain good agreement between the SEOBNR\_T and the NSBH-dedicated models with ${\cal{F}} > 0.99$,
and find no irregular behavior in the entire parameter space.
We also find that the IMRPhenomP\_T waveform is very consistent with the SEOBNR\_T waveform.
Additionally, we have verified that the correlation between any two of the $M_c, \eta,$ and $\chi_{\rm BH}$ parameters
suddenly decreased significantly or changed the sign in the bottom-left corner for the NSBH-dedicated models,
and these irregular behaviors were independent of the accuracy of the matrix inversion. 
Therefore, we conclude that the issue is with the Fisher Matrix method and not using a particular
 waveform model which is otherwise sound, when combined with Fisher matrix studies,  
 that makes no sense at all.

\subsection{Single detector analysis}

\begin{table}[b]
\centering
\begin{tabular}{c cccc}
\hline\hline
Population & Parameter space &$\chi_{\rm BH} $& $\deff$ &  $d_L$      \\
\hline
Pop-I       &  $m_1$--$m_2$    &0&  $40{\rm Mpc}$ &   $\cdot \cdot \cdot$  \\
Pop-II       &  $m_1$--$m_2$--$\chi_{\rm BH} $ &$\cdot \cdot \cdot$  &  $40{\rm Mpc}$ &  $\cdot \cdot \cdot$\\
Pop-III       &  $m_1$--$m_2$--$\chi_{\rm BH} $--RA--DEC &$\cdot \cdot \cdot$  &  $\cdot \cdot \cdot$& $40{\rm Mpc}$\\
\hline\hline
\end{tabular}
\caption{Description of Pop--I, Pop--II, and Pop--III.
In Pop--III, we assume the optimal binary orientation ($\theta_{JN}=\Psi=0$).}
\label{tab.pops}
\end{table}

In this subsection, we perform a single-detector analysis using the aLIGO and the CE PSDs, respectively.
We prepare three populations of NSBH sources.
In the first population (Pop-I), we produce $10^3$ Monte Carlo sources distributed in the 2--D $m_1$--$m_2$ space assuming $\chi_{\rm BH}=0$ and $\deff=40{\rm Mpc}$.
In the second population (Pop-II), we produce $2 \times 10^3$ Monte Carlo sources distributed in the 3--D $m_1$--$m_2$-$\chi_{\rm BH}$ space assuming $\deff=40{\rm Mpc}$.
Finally, in the third population (Pop-III), we produce $3 \times 10^3$ Monte Carlo sources distributed in the 5--D $m_1$--$m_2$-$\chi_{\rm BH}$--RA-DEC space assuming $d_L=40{\rm Mpc}$ and $\theta_{JN}=\Psi=0$.
The ranges of the intrinsic parameters are given as $[4\msun, 10\msun]$ for $m_1$, $[1\msun, 2\msun]$ for $m_2$, and $[-0.9,0.9]$ for $\chi_{\rm BH}$.
We do not restrict the range of the sky position, i.e., $[0,2\pi]$ for RA and $[0,\pi]$ for DEC.
The description of the populations is summarized in Table \ref{tab.pops}.
Using these populations, we calculate the SNRs and obtain the measurement errors from the $6\times 6$ Fisher matrices with the variables $\{M_c,\eta,\chi_{\rm BH},\lambda_{\rm NS}, t_c,\phi_c\}$.

\begin{figure}[t]
\begin{center}
\includegraphics[width=\columnwidth]{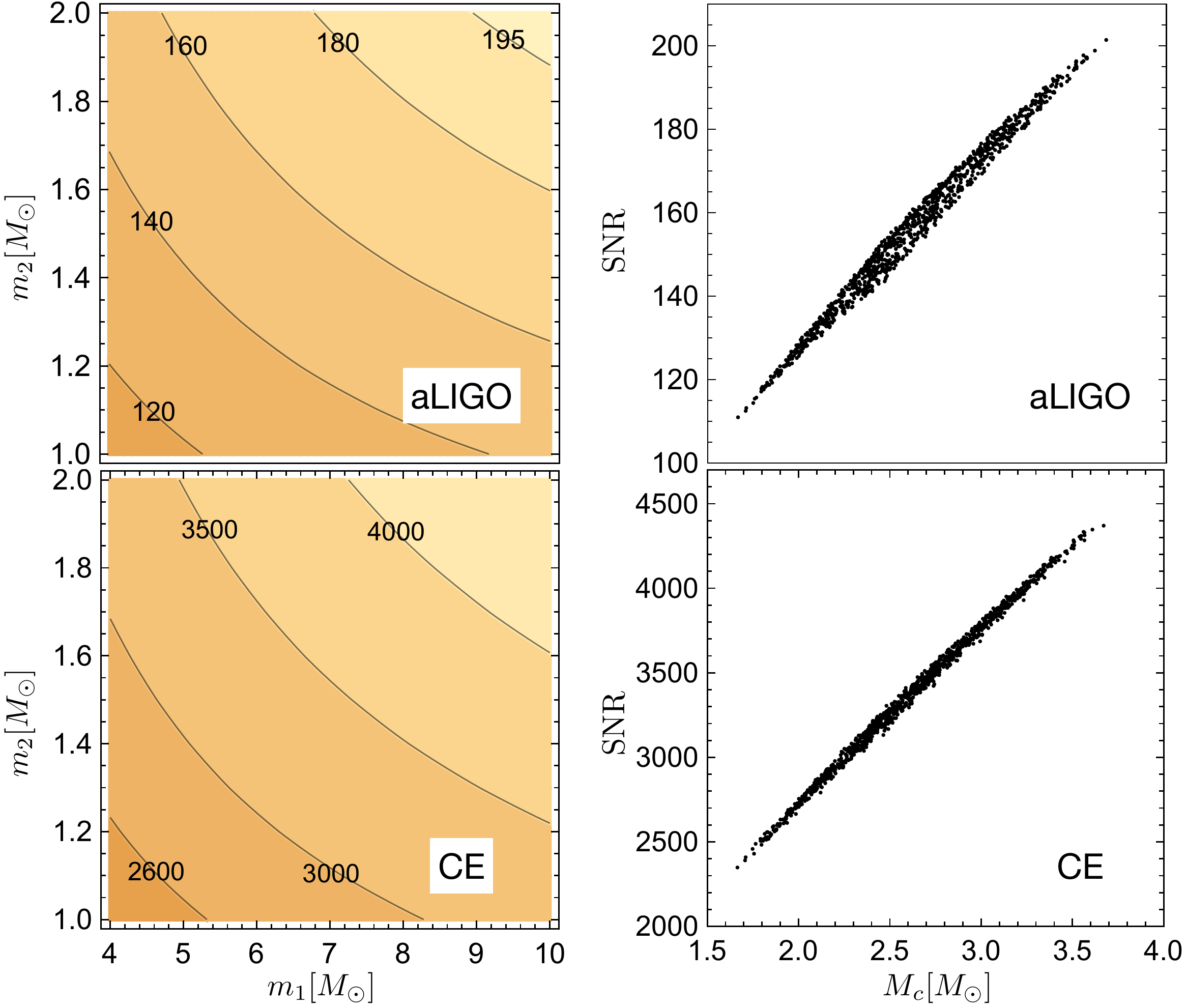}
\caption{\label{fig.SNR-comparison} SNRs for the sources in Pop-I distributed in the $m_1$--$m_2$ plane (left)
and given as a function of the chirp mass (right).}
\end{center}
\end{figure}

Figure \ref{fig.SNR-comparison} shows the SNR results for the sources in Pop-I.
The left panel displays the SNRs in the $m_1$--$m_2$ plane for the aLIGO and the CE detectors.
For both detectors, the contours show a very smooth and clear trend.
The right panel shows the SNRs as a function of the chirp mass ($M_c$),
and this result clearly describes the SNR's strong dependence on $M_c$.
The CE detector can have much larger SNRs than those for aLIGO,
and the SNR ratios ($ \rho_{\rm CE}/\rho_{\rm aLIGO}$) are distributed in a very narrow range $\sim [21.4, 22.2]$.

\begin{figure}[t]
\begin{center}
\includegraphics[width=\columnwidth]{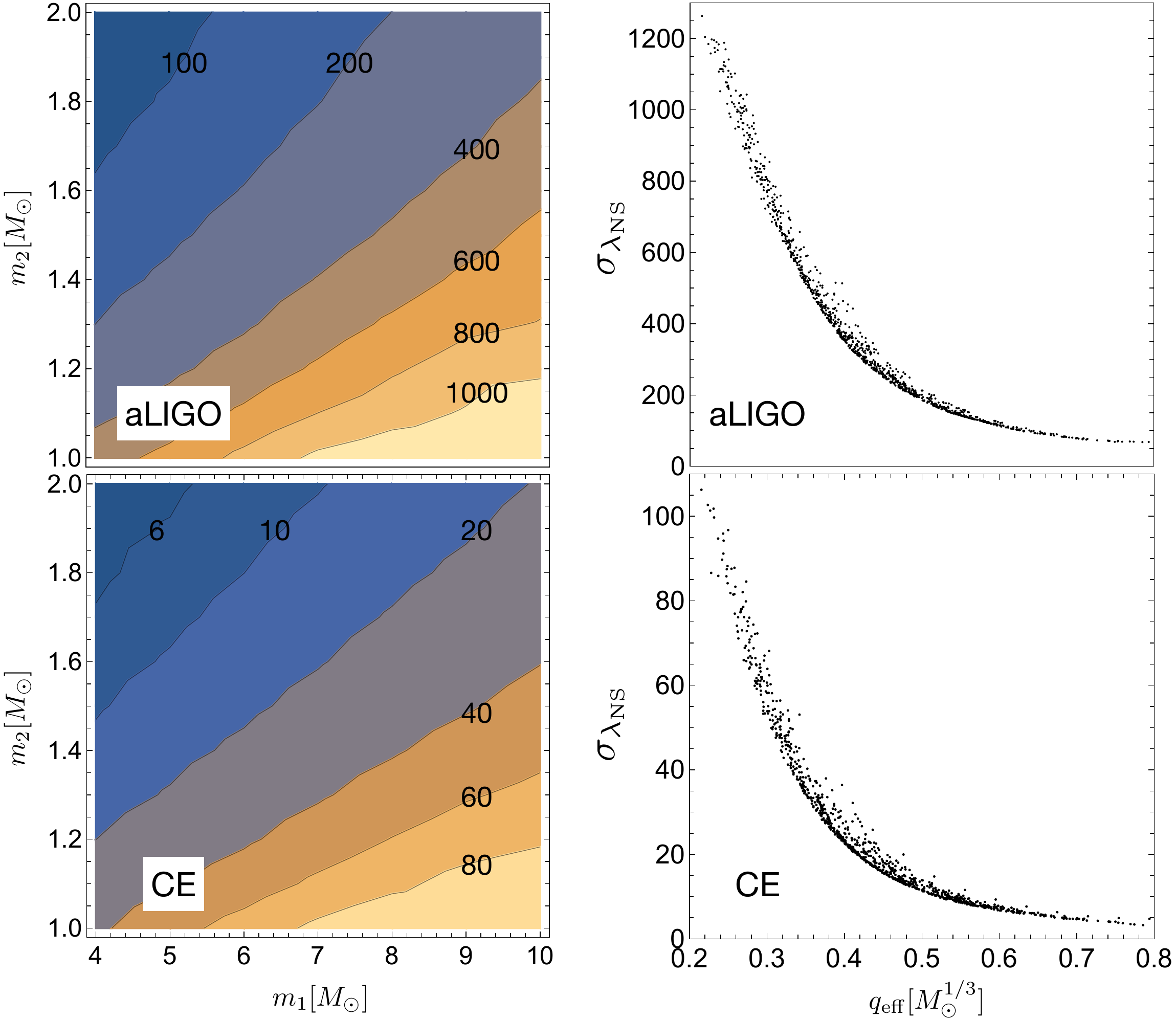}
\caption{\label{fig.error-lambda-comparison} Measurement errors ($\sigma_{\lambda_{\rm NS}}$) for the sources in Pop-I 
distributed in the $m_1$--$m_2$ plane (left)
and given as a function of the effective mass ratio $q_{\rm eff} \equiv m_2/m_1^{2/3}$ (right).}
\end{center}
\end{figure}

Similarly, Figure \ref{fig.error-lambda-comparison} shows the measurement errors of the NS tidal deformability for the sources in Pop-I.
As in the case of SNR, the error contours also show similar trends between aLIGO and CE.
However, the errors seem to depend mainly on the mass ratio rather than the chirp mass.
We have verified that the error distribution exhibits the narrowest band if we display the errors as a function of the effective mass
ratio defined by $q_{\rm eff} \equiv m_2/m_1^{2/3}$, which is shown in the right panel.
We performed the same analysis using the TaylorF2 waveform model,
and the errors showed a narrower distribution like a thin curve (see Appendix \ref{ap.TaylorF2}).
We tried to figure out how $q_{\rm eff}$ could be derived from the post-Newtonian equation
through the Fisher matrix formalism, but it was unsuccessful.
The definition of $q_{\rm eff}$ was empirically chosen from the shape of the contours.
Further study is needed for a reasonable explanation, and we leave it for future work.
The CE detector can measure the tidal deformability much more accurately than aLIGO,
and the error ratios ($ \sigma_{\rm CE}/\sigma_{\rm aLIGO}$) are distributed in a range $\sim [5.5, 8.3]\%$.

\begin{figure}[t]
\begin{center}
\includegraphics[width=\columnwidth]{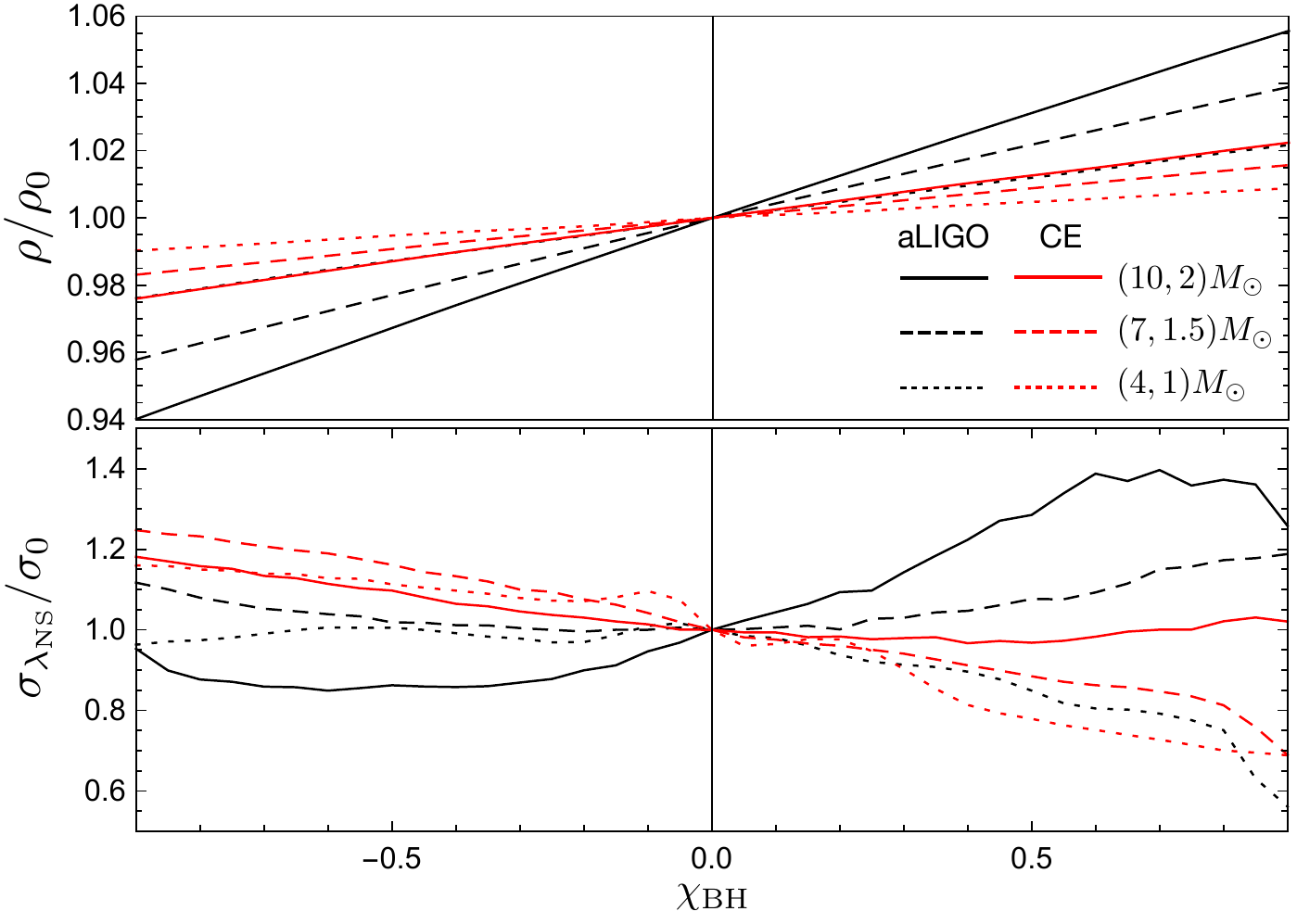}
\caption{\label{fig.chi-dependence} Dependence of the SNR ($\rho$) (upper panel) and the error ($\sigma_{\lambda_{\rm NS}}$) (lower panel) on the BH spin. Here, $\rho_0$ and $\sigma_0$ denote the values of the SNR 
and the error at $\chi_{\rm BH}=0$, respectively.
We assume $\deff=40{\rm Mpc}$.}
\end{center}
\end{figure}

To see the dependence on the BH spin ($\chi_{\rm BH}$), we select three sources with different masses, 
and for these signals, we calculate the SNRs and the errors $\sigma_{\lambda_{\rm NS}}$ varying the true value of $\chi_{\rm BH}$ in the range $[-0.9, 0.9]$.
The results are given in Fig. \ref{fig.chi-dependence}.
For efficiency, we give the fractional values, where $\rho_0$ and $\sigma_0$ denote the values of the SNR and the error at $\chi_{\rm BH}=0$, respectively.
In the upper panel, the SNR is quite symmetric between positive and negative spins
and depends almost linearly on the BH spin for all sources.
The variation in $\rho/\rho_0$ is more pronounced for aLIGO
and larger for more massive binaries.
The largest variation is $\sim 6\%$ and $\sim 2.6\%$ at $\chi_{\rm BH}=-0.9$ for aLIGO and CE, respectively.
In the lower panel, the errors do not show consistent behavior between aLIGO and CE
as well as between the sources.
The variation in $\sigma_{\lambda_{\rm NS}}/\sigma_0$ is much larger than the variation in $\rho/\rho_0$
and increases up to $\sim \pm 40\%$ in our spin range for both detectors.

\begin{figure}[t]
\begin{center}
\includegraphics[width=\columnwidth]{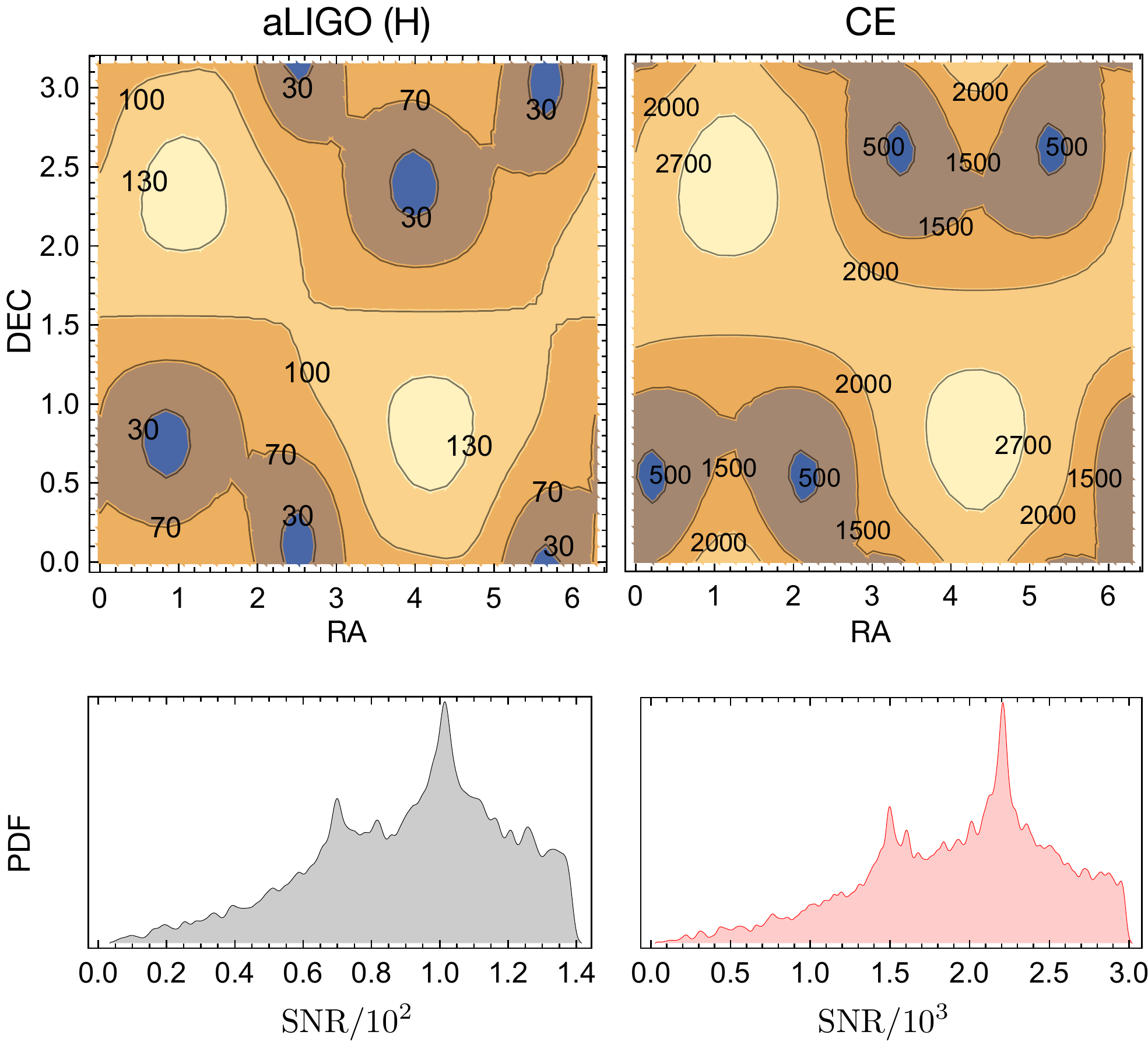}
\caption{\label{fig.ra-dec-snr} SNR distribution in the RA--DEC (sky position) plane  obtained from a single-detector analysis for the fiducial NSBH source ($m_1=5\msun, m_2=1.4\msun, \chi_{\rm BH}=0$) assuming $d_L=40{\rm Mpc}$ and $\theta_{JN}=\Psi=0$.}
\end{center}
\end{figure}

Geometrically, an L--Shaped single detector configuration such as LIGO can have two optimal positions and four hidden positions.
The optimal positions correspond to the two directions perpendicular to the detector-arms plane,
and the hidden positions correspond to the four directions that bisect the detector-arms axes. 
Figure \ref{fig.ra-dec-snr} shows the SNR distribution in the RA--DEC (sky position) plane 
obtained by using the aLIGO-Hanford (H) and the CE detectors, respectively.
Again, we assume the fiducial NSBH source ($m_1=5\msun, m_2=1.4\msun, \chi_{\rm BH}=0$)
with a fixed distance ($d_L=40{\rm Mpc}$) and the optimal orbital orientation ($\theta_{JN}=\Psi=0$).
In the upper panel, the optimal sky positions, where the SNR is maximum, for H and CE are very similar
because they are located at a similar site.
However, they have different hidden sky positions due to their different orientations.
On the other hand, the shape of the PDF curve is almost identical between the two detectors 
independently of the detector location 
because the two detectors have the same L-shaped arms and the SNRs are averaged over the entire sky position.
The SNR PDF gradually increases in the low region but falls quickly at the end. 
The SNR value ranges from zero to $\sim 140 (3000)$ and is maximum at $\sim 100 (2200)$ for aLIGO (CE).

\begin{figure}[t]
\begin{center}
\includegraphics[width=\columnwidth]{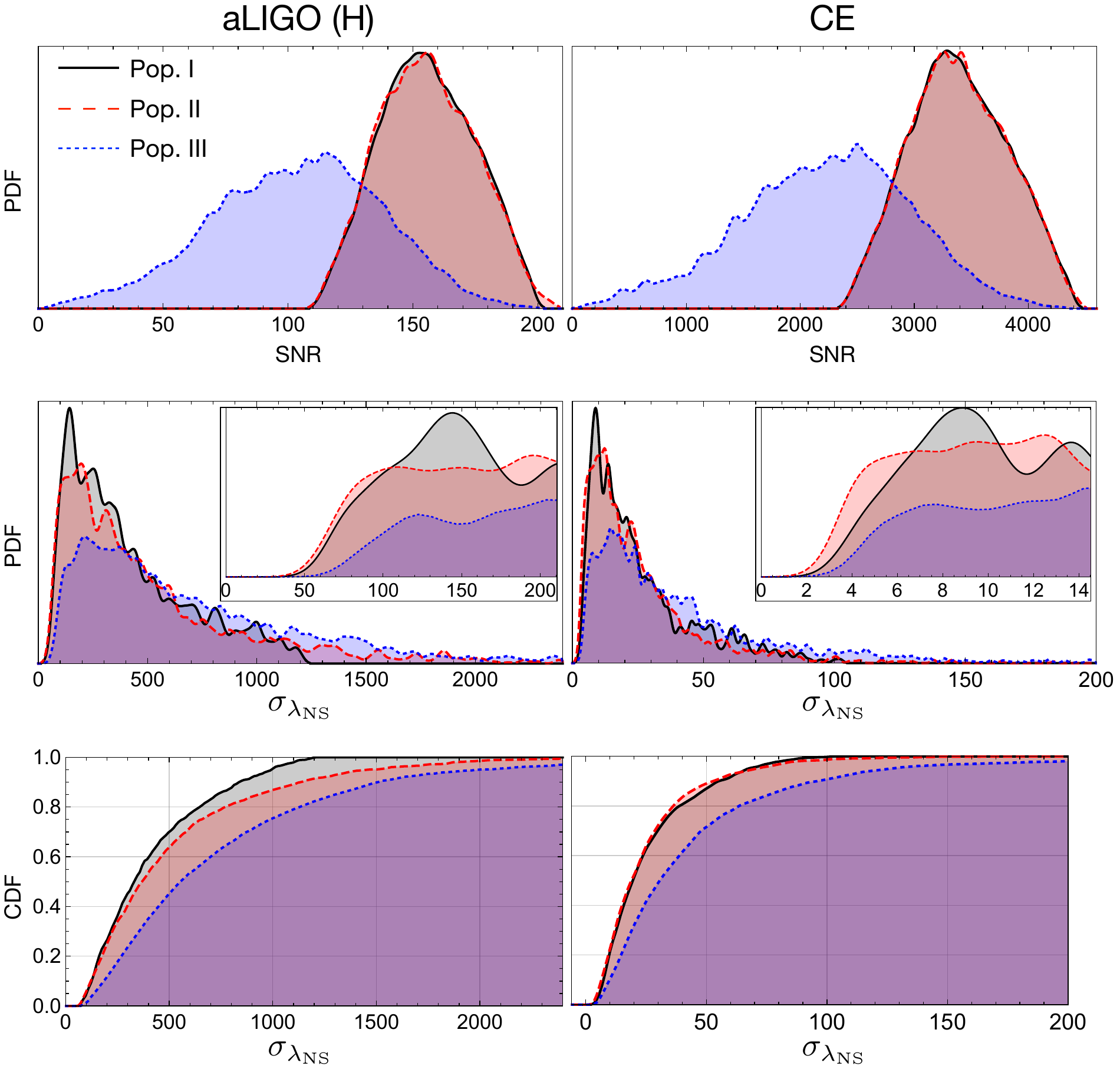}
\caption{\label{fig.pop123-snr-error} PDF of the SNR (top panel),  PDF of $\sigma_{\lambda_{\rm NS}}$ (middle), 
and CDF of $\sigma_{\lambda_{\rm NS}}$ (bottom) for the sources in Pop-I, Pop-II, and Pop-III.
The zoom-in plot shows the PDFs in the low region.}
\end{center}
\end{figure}

To obtain generalized PDFs of the SNR and $\sigma_{\lambda_{\rm NS}}$,
we perform the Fisher matrix analysis on the sources in Pop-I, Pop-II, and Pop-III, respectively.
In Fig. \ref{fig.pop123-snr-error}, we display the PDF of the SNR (top),
the PDF of $\sigma_{\lambda_{\rm NS}}$ (middle), and its cumulative distribution function (CDF) (bottom).
In the top panel, it can be seen that
although the parameter space has been extended from two to three dimensions by including the BH spin parameter,
there is no noticeable difference between the  Pop-I and Pop-II PDFs.
This is because the SNR fluctuations due to the BH spin are only a few percent as shown in Fig. \ref{fig.chi-dependence},
and thus the overall distribution is rarely affected by the spin.
However, the PDF of Pop-III is more widely distributed in the lower region. 
In this case, the SNR can have a very small value around the hidden positions.
The SNR ranges from $\sim 110 (2200)$ to $\sim 210 (4500)$ for aLIGO (CE) in both the Pop-I and the Pop-II cases.
In the case of Pop-III, the SNR value starts from zero but can increase to the same value as the maximum value of Pop-II
when the source is located at the optimal sky position.
In the middle panel,
the PDF curves of $\sigma_{\lambda_{\rm NS}}$ are similar in shape between H and CE but different between the three populations.
In the case of Pop-I, the PDF has a sharp peak in the low region and slowly decreases.
As the parameter dimension of the population increases, the peak becomes lower and flatter
and the tail becomes longer.
The errors range broadly from $\sim 40 (2)$ to more than $2500 (200)$ for H (CE),
but most errors are concentrated in the low region.
The results of $\sigma_{\lambda_{\rm NS}}$ for the single-detector analysis are summarized in Table \ref{tab.errors}.
In the case of H, the values of $\sigma^{\rm peak}$, where the PDF is maximum,
are given as $\sigma^{\rm peak} \sim 140, 190,$ and 210,
and the values of $\sigma^{90\%}$, where ${\rm CDF}=0.9$, are given as $\sigma^{90\%} \sim 840, \sim 1130$, and  $\sim1500$ for Pop-I, Pop-II, and Pop-III, respectively,
In the case of CE, those are $\sigma^{\rm peak} \sim 9, 14,$ and 15
and $\sigma^{90\%} \sim 55, \sim 56$, and  $\sim 96$, respectively.
Overall, our results suggest an 
improvement of about 20 times between H and CE in tidal deformability measurements.
On the other hand, 
Lackey \etal \cite{PhysRevD.89.043009} also estimated the measurement errors of $\lambda_{\rm NS}$ from NSBH systems for single aLIGO and (L-shaped) ET detectors
and found an order of magnitude better accuracy in the case of ET.  
The PDFs for the other three intrinsic parameters are given in Appendix \ref{ap.other-results}.

\begin{table}[b]
\centering
\begin{tabular}{c cccccccc}
\hline\hline
 &  \multicolumn{4}{c}{aLIGO (H) }& \multicolumn{4}{c}{CE}\\
 \cmidrule(lr){2-5}\cmidrule(lr){6-9}
Population   &    $\sigma^{\rm peak}$ &  $\sigma^{30\%}$ &   $\sigma^{60\%}$ &  $\sigma^{90\%}$ &      $\sigma^{\rm peak}$ &  $\sigma^{30\%}$ &   $\sigma^{60\%}$ &  $\sigma^{90\%}$ \\
\hline
Pop-I& 140&   220    & 400&   840&9&  13     &  24&55   \\
Pop-II&190&   225   & 460&   1130     &14&12&23&56  \\
Pop-III& 210&  360 & 700& 1500 &  15&19&39&96   \\
\hline\hline
\end{tabular}
\caption{Summary of the measurement errors for the single-detector analysis. Here, $\sigma^{\rm peak}$ indicates the value where the PDF is maximum,
and $\sigma^{{\rm n}\%}$ corresponds to the value where ${\rm CDF}={\rm n}\%$.}
\label{tab.errors}
\end{table}

\subsection{Multidetector network analysis}

In this subsection, we perform a multi-detector analysis using the networks of the 2G and the 3G detectors, respectively.
We adopt the 2G network HLVK consisting of aLIGO-Hanford (H), aLIGO-Livingstone (L), advanced Virgo (V), and KAGRA (K), 
and the 3G network ETCE consisting of Einstein Telescope (ET) and Cosmic Explorer (CE).
In a network, the SNR can be given by
\be
\rho_{\rm net}^2=\sum_i \rho_{i}^2,
\ee
where $\rho_i$ indicates the SNR of the $i$--th detector. 
Similarly, the measurement error can be calculated from the network Fisher matrix given by
\be
\Gamma_{\rm net}=\sum_i  \Gamma_i.
\ee

\begin{figure}[t]
\begin{center}
\includegraphics[width=\columnwidth]{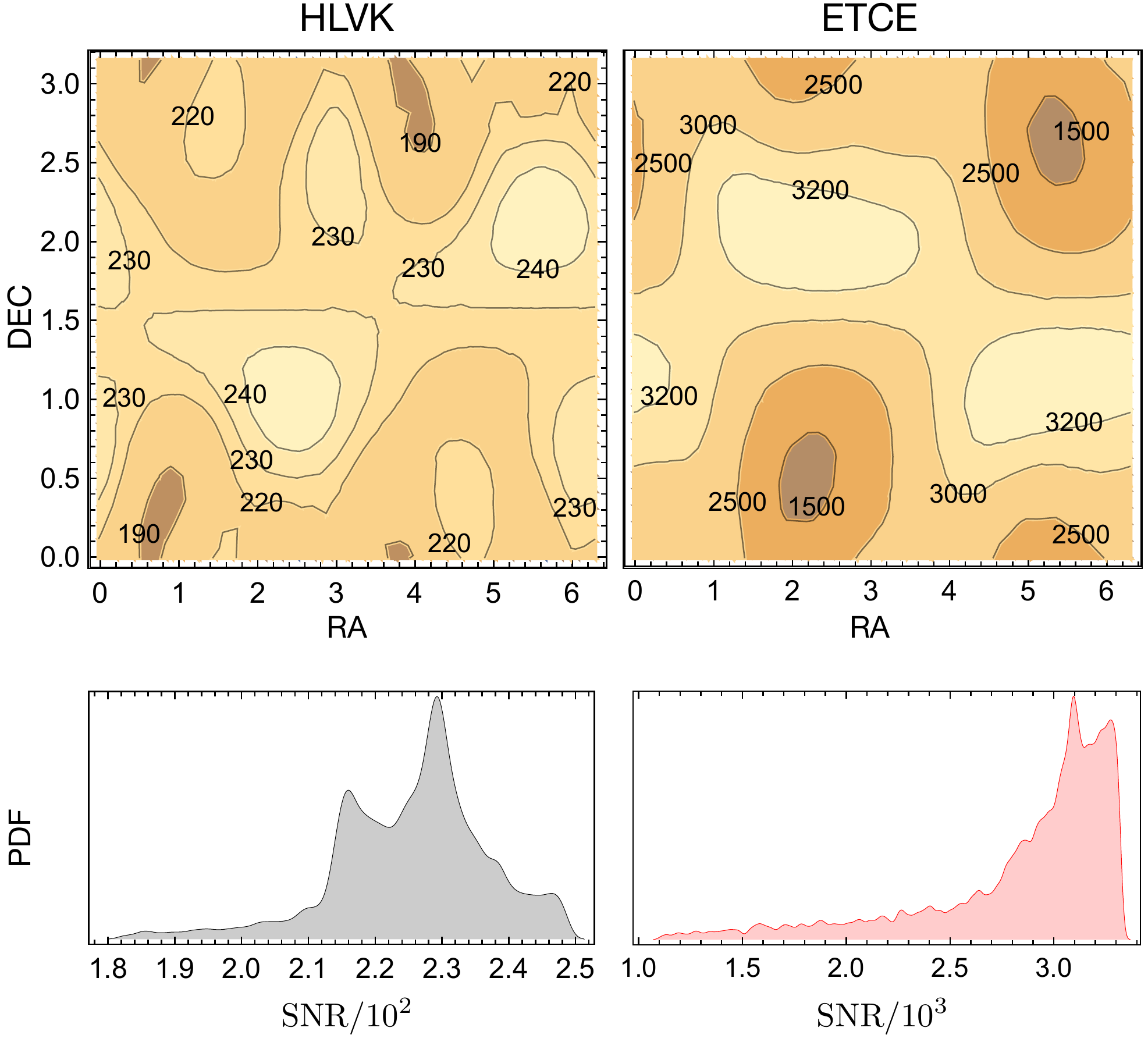}
\caption{\label{fig.ra-dec-network-snr} SNR distribution in the RA--DEC plane obtained from 
a multi-detector analysis for the fiducial NSBH source ($m_1=5\msun, m_2=1.4\msun, \chi_{\rm BH}=0$) 
assuming $d_L=40{\rm Mpc}$ and $\theta_{JN}=\Psi=0$.}
\end{center}
\end{figure}

Figure \ref{fig.ra-dec-network-snr} shows the SNR distribution in the RA--DEC plane
obtained by using the 2G and the 3G networks, respectively.
We assume the fiducial NSBH source ($m_1=5\msun, m_2=1.4\msun, \chi_{\rm BH}=0$)
with a fixed distance ($d_L=40{\rm Mpc}$) and the optimal orbital orientation.
Unlike the single detector case, the trends of the SNR contours are quite 
different between the 2G and the 3G networks, and there are no hidden positions.
The SNR ranges from $\sim 180(1100)$ to $\sim 250(3400)$ for the 2G(3G) network,
and the shapes of the PDF curves are significantly different between the two networks.
The PDF is maximum at $\rho \sim 230$ and $\sim 3100$ for the 2G and the 3G networks, respectively.

\begin{figure}[t]
\begin{center}
\includegraphics[width=\columnwidth]{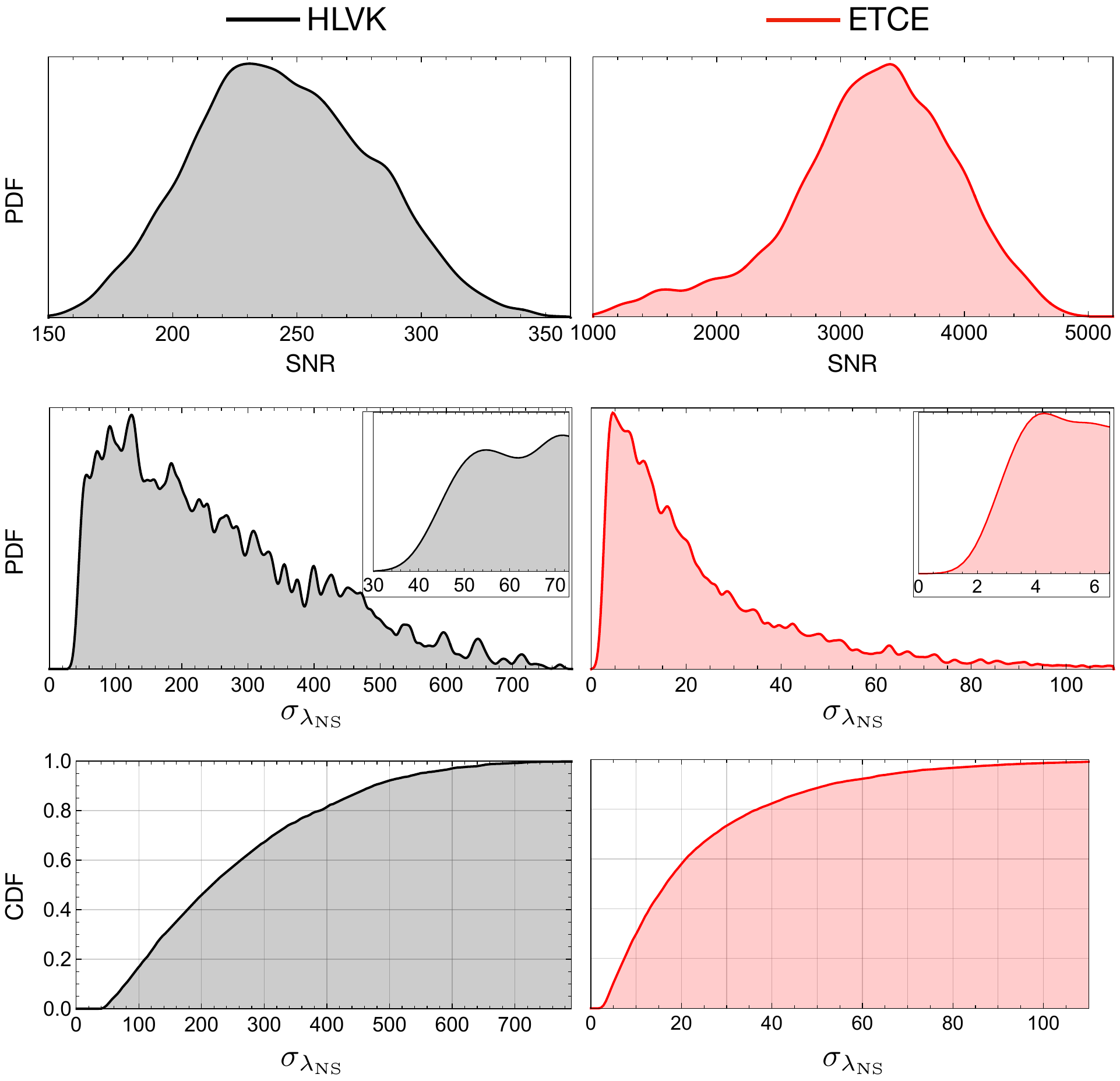}
\caption{\label{fig.5-d-histogram} PDF of the SNR (top panel),  PDF of $\sigma_{\lambda_{\rm NS}}$ (middle), 
and CDF of $\sigma_{\lambda_{\rm NS}}$ (bottom) for the sources in Pop-III.
The zoom-in plot shows the PDFs in the low region.}
\end{center}
\end{figure}

In Fig. \ref{fig.5-d-histogram}, we display the PDF of the SNR (top),
the PDF of $\sigma_{\lambda_{\rm NS}}$ (middle), and its CDF (bottom) for the sources in Pop-III.
In the top panel, the SNR ranges from $\sim 150$ to $\sim 360$ for the 2G network 
and from $\sim 1000$ to $\sim 5000$ for the 3G network.
The SNR PDFs are maximum at $\sim 230$ and $\sim 3400$ for the 2G and the 3G networks, respectively,
and they are much larger than those in the single detector case.
In the middle and the bottom panels, 
the overall trends of the PDF and the CDF curves are similar to those in the single detector case,
but of the two networks, the 3G network can have a PDF curve with a sharper peak and a longer tail between the two networks.
The results of $\sigma_{\lambda_{\rm NS}}$ for the multi-detector analysis are summarized in Table \ref{tab.network-errors}.
The  $\sigma^{\rm peak}$ values are given as $\sigma^{\rm peak} \sim 130$ and $\sim 4$
and the $\sigma^{90\%}$ values are given as $\sigma^{90\%} \sim 470$ and  $\sim 53$
for the 2G and the 3G networks, respectively.
The PDFs for the other three intrinsic parameters are given in Appendix \ref{ap.other-results}.

\begin{table}[b]
\centering
\begin{tabular}{c cccccccc}
\hline\hline
 &  \multicolumn{4}{c}{HLVK }& \multicolumn{4}{c}{ETCE}\\
  \cmidrule(lr){2-5}\cmidrule(lr){6-9}
Population   &    $\sigma^{\rm peak}$ &  $\sigma^{30\%}$ &   $\sigma^{60\%}$ &  $\sigma^{90\%}$ &      $\sigma^{\rm peak}$ &  $\sigma^{30\%}$ &   $\sigma^{60\%}$ &  $\sigma^{90\%}$ \\\hline
Pop-III& 130&   140    & 260&   470 & 4&   10   & 21&   53  \\
\hline\hline
\end{tabular}
\caption{The same as in Table \ref{tab.errors} but obtained with the multi-detector networks.}
\label{tab.network-errors}
\end{table}

\subsection{Constraining EOS models}

\begin{figure}[t]
\begin{center}
\includegraphics[width=\columnwidth]{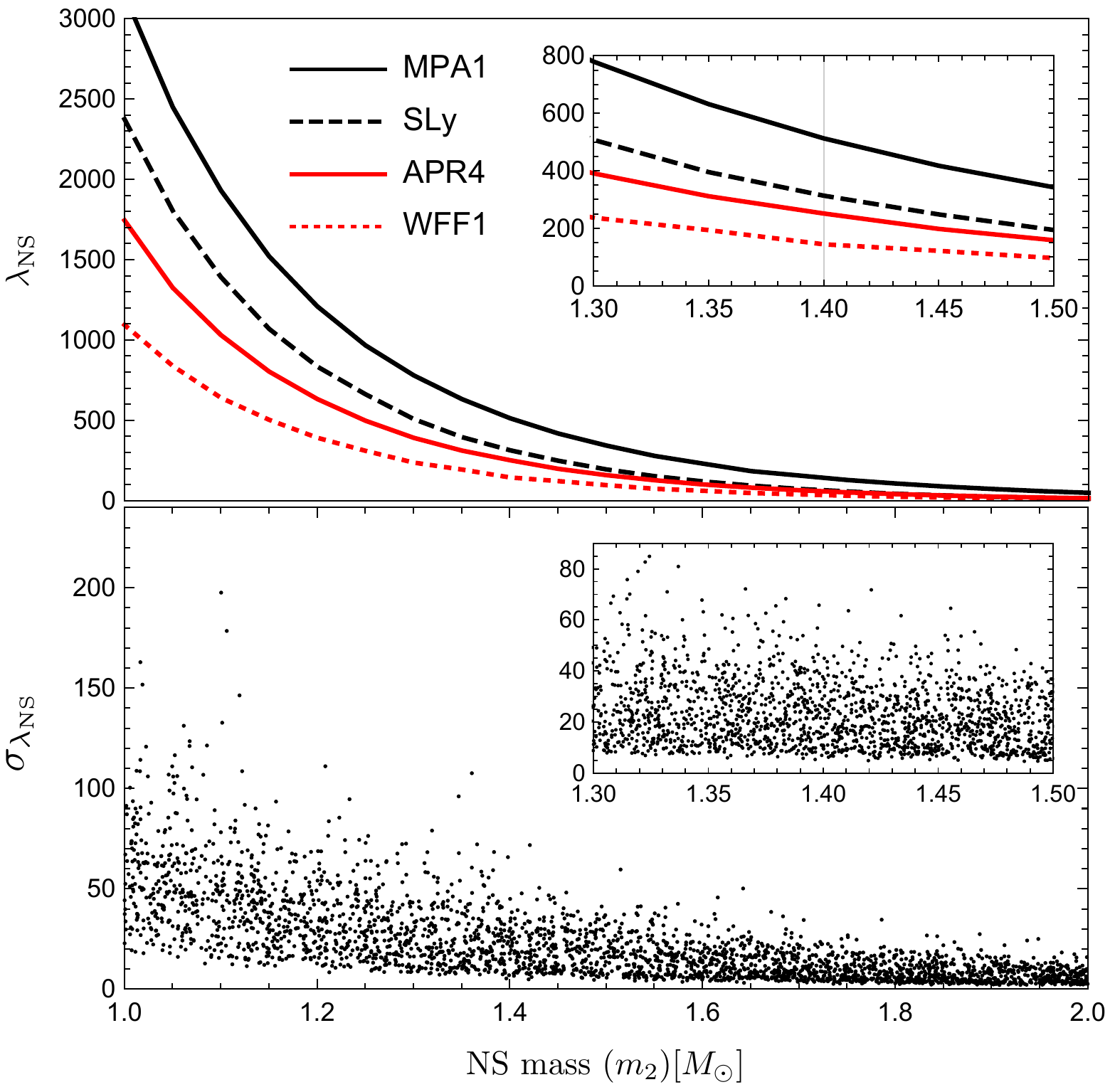}
\caption{\label{fig.EOS-curves}Upper panel : Tidal parameter value given as a function of the NS mass for different EOS models.
Lower panel : Distribution of $\sigma_{\lambda_{\rm NS}}$ for the sources in Pop-III obtained with the 3G network.}
\end{center}
\end{figure}

Here, we utilize our results to evaluate how well the theoretical EOS models can be constrained by GW parameter estimation
for NSBH signals.
We choose four EOS models, WFF1, APR4, SLy, and MPA1 \cite{PhysRevD.79.124032}, 
which lie well inside of the $90\%$ credible region of the PDFs 
for all waveform models used in the analysis of GW170817 \cite{GW170817PE}.
In Fig. \ref{fig.EOS-curves}, the upper panel presents the tidal parameter $\lambda_{\rm NS}$
as a function of the NS mass ($m_2$).
For all models, $\lambda_{\rm NS}$ monotonically decreases with increasing NS mass,
and the stiffer EOS model gives a higher parameter value for a given NS mass.
The parameter values of $\lambda_{\rm NS}$ at $m_2=1\msun$ are widely distributed from $\sim 1000$ to $\sim 3000$
depending on the EOS model,
but they are less than 50 at $m_2=2\msun$ for all models.
The lower panel displays the distribution of $\sigma_{\lambda_{\rm NS}}$ 
for the sources in Pop-III obtained with the 3G network.
The error distribution is also wider for smaller NS masses.
However, the distribution width is much narrower than that of $\lambda_{\rm NS}$ (in the upper panel) in the low-mass region
but is comparable in the high-mass region.
This implies that the smaller the NS mass, the better the recovered value of $\lambda_{\rm NS}$ can be constrained
between different EOS models.

\begin{figure}[t]
\begin{center}
\includegraphics[width=\columnwidth]{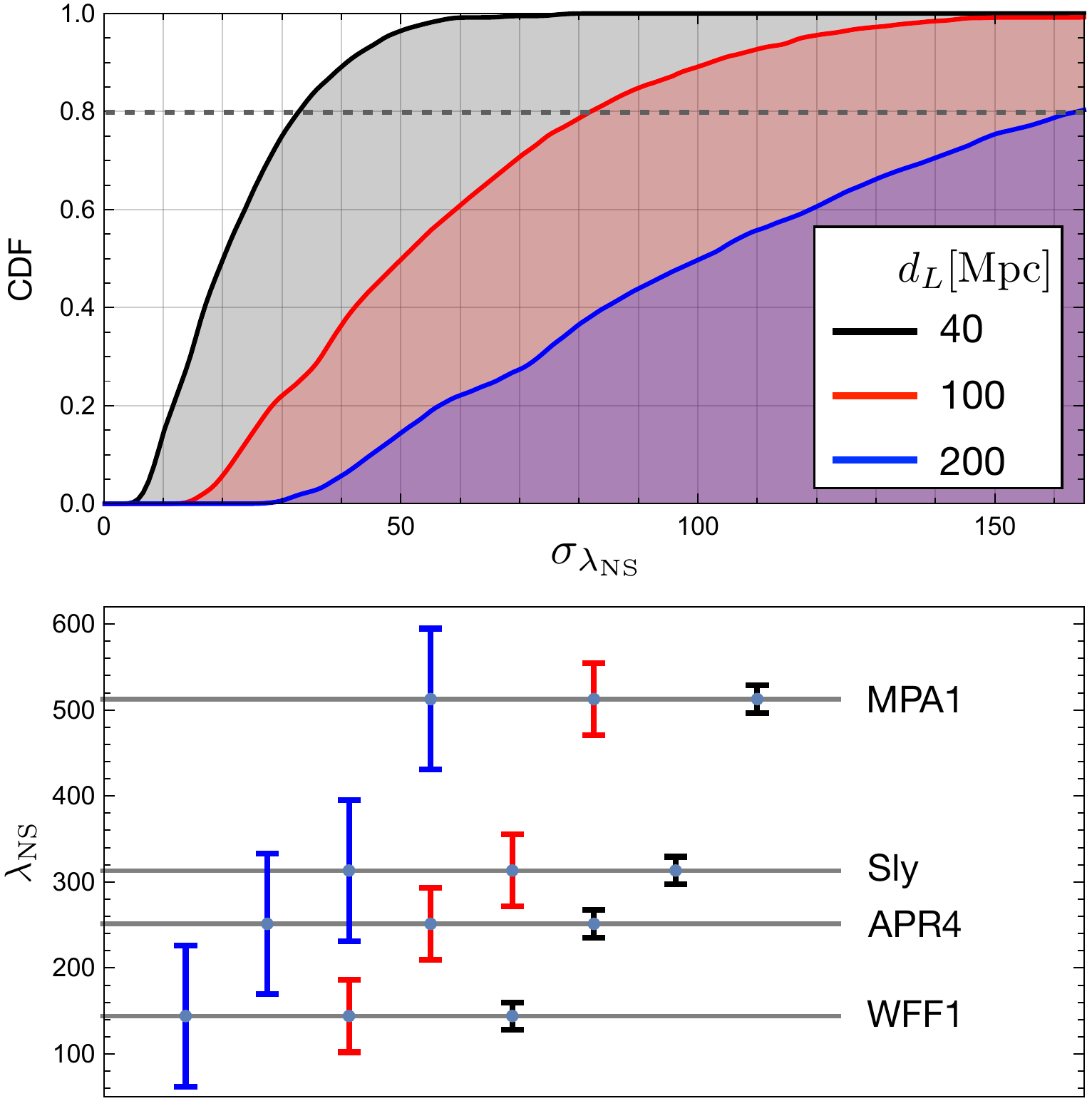}
\caption{\label{fig.EOS-measuring} Upper panel : CDF curves of $\sigma_{\lambda_{\rm NS}}$ obtained by using $10^3$ Monte Carlo sources 
in Pop-IV. We assume three different distances,
$d_L=40, 100$, and $200{\rm Mpc}$, and use the 3G network. 
We find $\sigma^{80\%} \sim 33, 82,$ and 164 for $d_L=40,100,$ and $200{\rm Mpc}$, respectively.
Lower panel : The value of $\lambda_{\rm NS}$ according to the EOS model (gray). The error bars correspond to 
$\sigma_{\lambda_{\rm NS}}=33$ (black), 82 (red), and 164 (blue), respectively.}
\end{center}
\end{figure}

We provide a concrete example to illustrate how well the EOS models can be distinguished from each other.
We prepare a specific population Pop-IV, where $10^3$ Monte Carlo sources are distributed 
in the 4--D $m_1$--$\chi_{\rm BH}$--RA--DEC space with the NS mass fixed to $m_2=1.4\msun$.
For all sources, we assume $d_L=40{\rm Mpc}$ and $\theta_{JN}=\Psi=0$.
Thus, this population is equivalent to Pop.III except that the NS mass is fixed.
For all sources, we obtain the errors $\sigma_{\lambda_{\rm NS}}$ from the 6--D Fisher matrices
using the 3G network.
In Fig. \ref{fig.EOS-measuring},
the upper panel shows the CDF curves of $\sigma_{\lambda_{\rm NS}}$
for the sources in Pop-IV located at $d_L=40, 100$, and $200{\rm Mpc}$, respectively.
Note that the results for $d_L=100$ and $200{\rm Mpc}$ can be obtained simply by applying a scale factor
as $\sigma_{n{\rm Mpc}}=\sigma_{40\rm Mpc} \times n/40$.
The horizontal dashed line indicates ${\rm CDF}=0.8$.
We find $\sigma^{80\%} \sim 33, 82,$ and 164 for the cases $d_L=40,100,$ and $200{\rm Mpc}$, respectively,
and that means $80\%$ of the sources in Pop-IV satisfy $\sigma_{\lambda_{\rm NS}} \lesssim 33, 82,$ and 164, respectively.
These error scales are illustrated in the lower panel,
where the gray line indicates the value of $\lambda_{\rm NS}$ according to the EOS model for an NS with $m_2=1.4\msun$,
and the error bars correspond to 
$\sigma_{\lambda_{\rm NS}}=33$ (black), 82 (red), and 164 (blue), respectively.
This plot shows the measurability of each EOS model for the sources in Pop-IV located at a certain distance.
For example, if enough sources in Pop-IV located at $d_L \simeq 200{\rm Mpc}$ are detected by the 3G network,
$\sim 20\%$ of their parameter estimation results cannot distinguish the two models SLy and APR4 at the 1--$\sigma$ level,
where we assume that the true value of $\lambda_{\rm NS}$ is determined by SLy or APR4.
On the contrary, if $d_L \simeq 100{\rm Mpc}$, $\sim 80\%$ of the parameter estimation results can distinguish between all EOS models well.

\subsection{Dependence on the EOS model}

\begin{figure}[t]
\begin{center}
\includegraphics[width=\columnwidth]{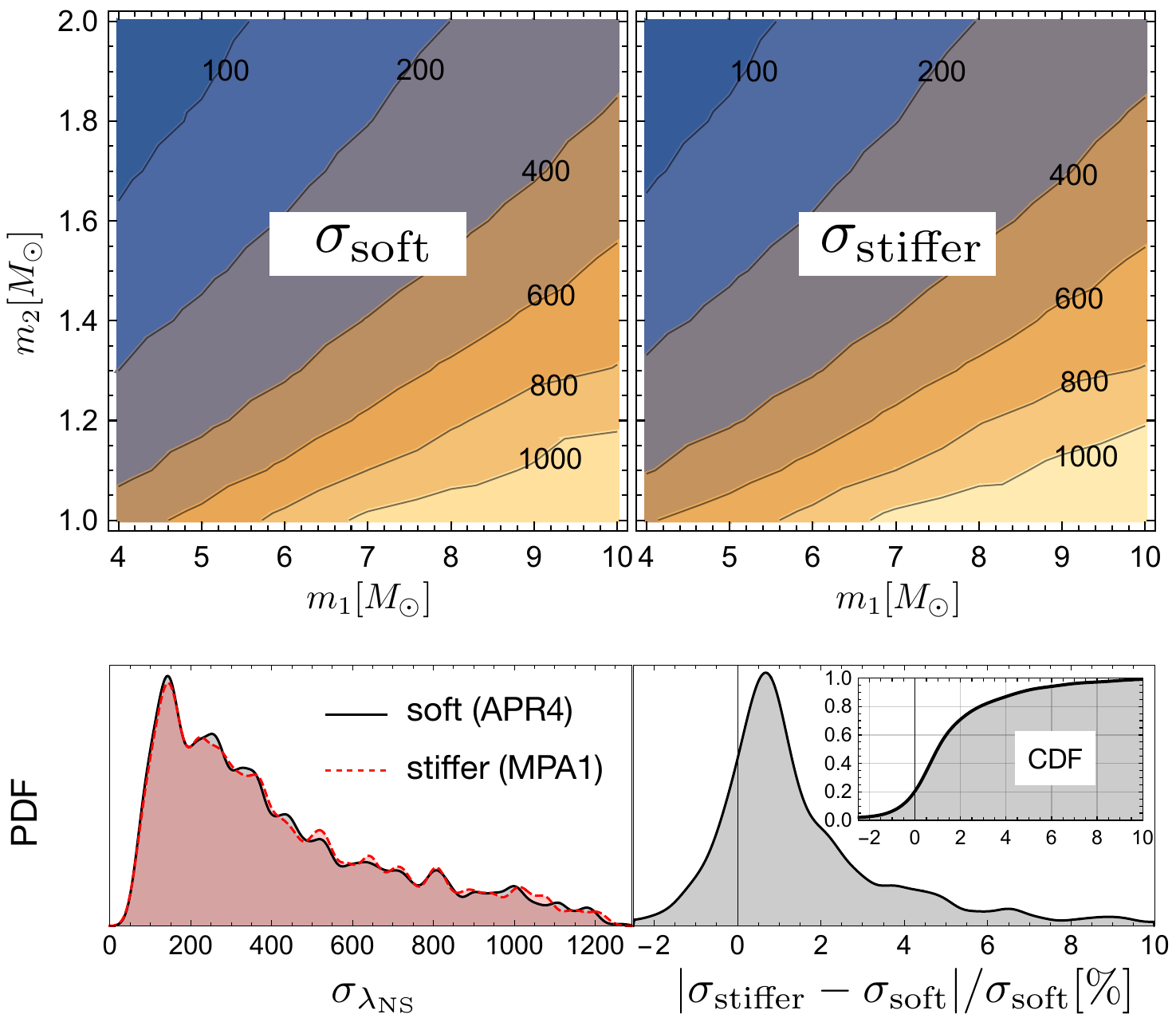}
\caption{\label{fig.soft-stiff-comparison}Upper panel : Contours of $\sigma_{\lambda_{\rm NS}}$ for the soft EOS model APR4 ($\sigma_{\rm soft}$) and 
the stiffer EOS model MPA1 ($\sigma_{\rm stiffer}$) obtained by using the sources in Pop-I with the aLIGO PSD.
Lower panel : PDFs of the error $\sigma_{\lambda_{\rm NS}}$ for the soft and the stiffer EOS models (left)
and the PDF and the CDF curves of the fractional difference $\Delta \sigma \equiv |\sigma_{\rm stiffer}-\sigma_{\rm soft}|/\sigma_{\rm soft}$.
Note that all sources satisfying $\Delta \sigma >7\%$ are concentrated in the bottom-left region ($m_1<5\msun, m_2 <1.4\msun$).}
\end{center}
\end{figure}

In the above analysis, we adopted the soft EOS model APR4 to choose the true value of $\lambda_{\rm NS}$.
In this subsection, we show how our results can be affected 
if the true value of $\lambda_{\rm NS}$ is given by other EOS models.
To this end, we perform the same Fisher matrix analysis for the sources in Pop-I with the aLIGO PSD
using the stiffer EOS model MPA1 \cite{PhysRevD.79.124032}.
The comparison result between APR4 and MPA1 is given in Fig. \ref{fig.soft-stiff-comparison}.
The upper panel shows the error contours for the soft ($\sigma_{\rm soft}$) and 
the stiffer ($\sigma_{\rm stiffer}$) EOS models in the $m_1$--$m_2$ plane,
where the result of $\sigma_{\rm soft}$ is taken from Fig. \ref{fig.error-lambda-comparison}.
The result exhibits a highly consistent contour trend between the soft and the stiffer EOS models
but shows small differences in the bottom-left region.
In the lower panel, the left plot displays the PDFs of $\sigma_{\lambda_{\rm NS}}$ for both EOS models.
For the same population of sources, the two EOS models provide PDF curves that are similar overall but slightly different.
These small differences are well quantified in the right plot, 
which shows the PDF and the CDF curves of the fractional difference $\Delta \sigma \equiv |\sigma_{\rm stiffer}-\sigma_{\rm soft}|/\sigma_{\rm soft}$
between the error for the stiffer ($\sigma_{\rm stiffer}$) and the error for the soft ($\sigma_{\rm soft}$) models.
We have verified that $\sim 90 (99)\%$ of the $10^3$ Monte Carlo sources can satisfy $\Delta \sigma <5(10)\%$,
and all of the sources with $\Delta \sigma >7\%$ are concentrated in the bottom-left region ($m_1<5\msun, m_2 <1.4\msun$).
Therefore, we conclude that the PDF curves obtained in this work are almost unaffected by the choice of the EOS model.

\section{Summary and discussion} \label{sec.summary}

In this work, we investigated how accurately the NS tidal deformability ($\lambda_{\rm NS}$) can be measured 
by GW parameter estimation for NSBH signals.
For the three populations of NSBH sources distributed in the 2--D $m_1$--$m_2$ space (Pop--I), the 3--D 
$m_1$--$m_2$--$\chi_{\rm BH}$ space (Pop--II), and the 5--D $m_1$--$m_2$--$\chi_{\rm BH}$--RA--DEC space (Pop--III),
we calculated the measurement errors ($\sigma_{\lambda_{\rm NS}}$) using the Fisher matrix method
and showed their general distributions as the 1--D PDFs.
We chose as our reference waveform model SEOBNR\_T, which is one of the recent IMR models including the NS tidal effect,
and adopted the four 2G detectors, H, L, V, and K,  and the two 3G detectors, ET and CE. 

We performed a single-detector analysis for the sources in Pop--I, Pop--II, and Pop--III
and compared the SNRs ($\rho$) and the measurement errors ($\sigma_{\lambda_{\rm NS}}$)
between the aLIGO and the CE detectors.
We find that the PDF curves of $\rho$ and $\sigma_{\lambda_{\rm NS}}$ are similar in shape between the two detectors,
but CE can achieve $\sim 15$ times better accuracy overall in the measurement of the NS tidal deformability $\lambda_{\rm NS}$.
We also performed a multi-detector analysis using the 2G and the 3G networks
for the sources in Pop--III.
The PDF curves of $\sigma_{\lambda_{\rm NS}}$ are maximum at $\sim 130$ and $\sim 4$ for the 2G and the 3G networks, respectively.
Overall, the 3G network can achieve $\sim 10$ times better accuracy in the measurement of $\lambda_{\rm NS}$.
The results for the single and the network detectors are summarized in Tables  \ref{tab.errors} and \ref{tab.network-errors}, respectively.
From the results for the $10^3$ Monte Carlo sources distributed in the 4--D $m_1$--$\chi_{\rm BH}$--RA--DEC space 
with the NS mass fixed to $m_2=1.4\msun$,
we found that if $d_L \simeq 100{\rm Mpc}$, the parameter estimation results for $\sim 80\%$ of the sources 
can distinguish between the EOS models, WFF1, APR4, SLy, and MPA1, at the 1--$\sigma$ level, using the 3G network.
Finally, we demonstrated that the PDF curves obtained in this work are almost independent of the true value
of $\lambda_{\rm NS}$.

To date, since there are only two NSBH signals, GW200105 and GW200115, detected by the LIGO-Virgo network,
it is difficult to predict how many NSBH signals will be detected in the 3G network era.
Our PDF curves may not be suitable for direct application to a small number of detection signals
because the PDFs represent the statistical distributions obtained by using Monte Carlo samples.

%%%%%%%%%%%%%%%%%%%%%%%%%%%%%%%%%%%%%%%%%%%%%%%%%%%%%%%%%%%

%=======	Acknowledgements ===========================	
%

\begin{acknowledgments}
This work was supported by the National Research
Foundation of Korea (NRF) grants funded by the Korea
government (No. 2018R1D1A1B07048599, No. 2019R1I1A2A01041244, and No. 2020R1C1C1003250)
\end{acknowledgments}

\section{RESULTS OF THE TaylorF2 WAVEFORM MODEL}  \label{ap.TaylorF2}

\begin{figure}[t]
\begin{center}
\includegraphics[width=\columnwidth]{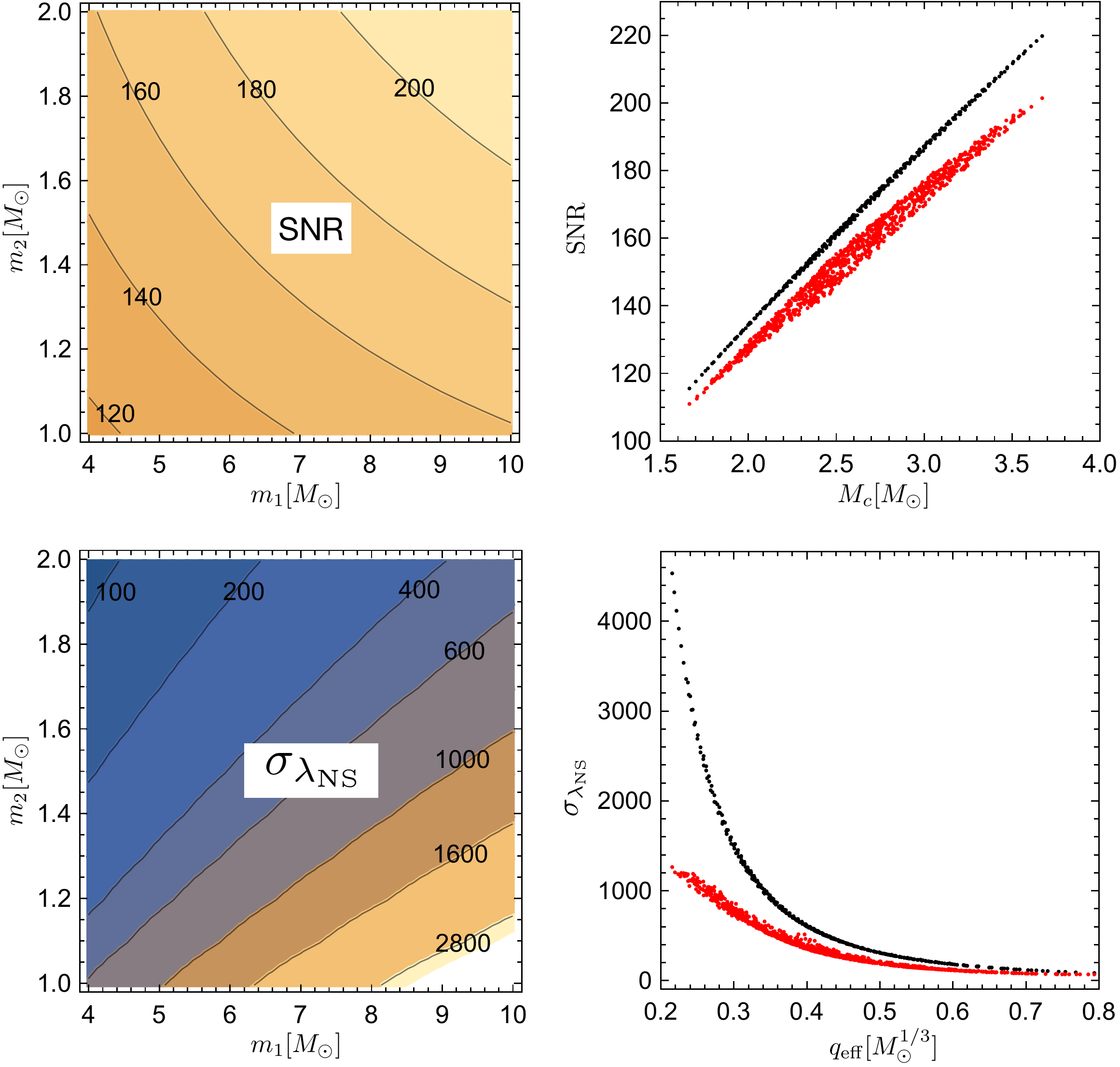}
\caption{\label{fig.TaylorF2} Upper panel : SNRs given in the $m_1$--$m_2$ plane (left) and given as a function of the chirp mass (right).  Lower panel : Measurement errors $\sigma_{\lambda_{\rm NS}}$ given in the $m_1$--$m_2$ plane (left) and given as a function of the effective mass ratio ($q_{\rm eff} \equiv m_2/m_1^{2/3}$).  These results are obtained by using the same sources (Pop-I) and the same PSD (aLIGO) as in Figs. \ref{fig.SNR-comparison} and \ref{fig.error-lambda-comparison}, but the TaylorF2 waveform model is used here.
For comparison, we also present the results for SEOBNR\_T (red).}
\end{center}
\end{figure}

Figure \ref{fig.TaylorF2} shows the SNRs and the measurement errors $\sigma_{\lambda_{\rm NS}}$ 
for the sources in Pop-I obtained by using the TaylorF2 waveform model.
We assume $f_{\rm max}$ to be the frequency at the innermost-stable-circular-orbit.
The upper (lower) panel clearly shows the strong dependence of the SNR (error) on the chirp mass (effective mass ratio).
These results are obtained by using the same sources (Pop-I) and the same PSD (aLIGO) as in Figs. \ref{fig.SNR-comparison} and \ref{fig.error-lambda-comparison}, hence directly comparable to the results for SEOBNR\_T.
Note that the results for $\sigma_{\lambda_{\rm NS}}$ are significantly different between TaylorF2 and SEOBNR\_T, 
especially in the low mass ratio region, indicating the inadequacy of the TaylorF2 model in our analysis.

\section{RESULTS OF THE OTHER INTRINSIC PARAMETERS}  \label{ap.other-results}

\begin{figure}[t]
\begin{center}
\includegraphics[width=\columnwidth]{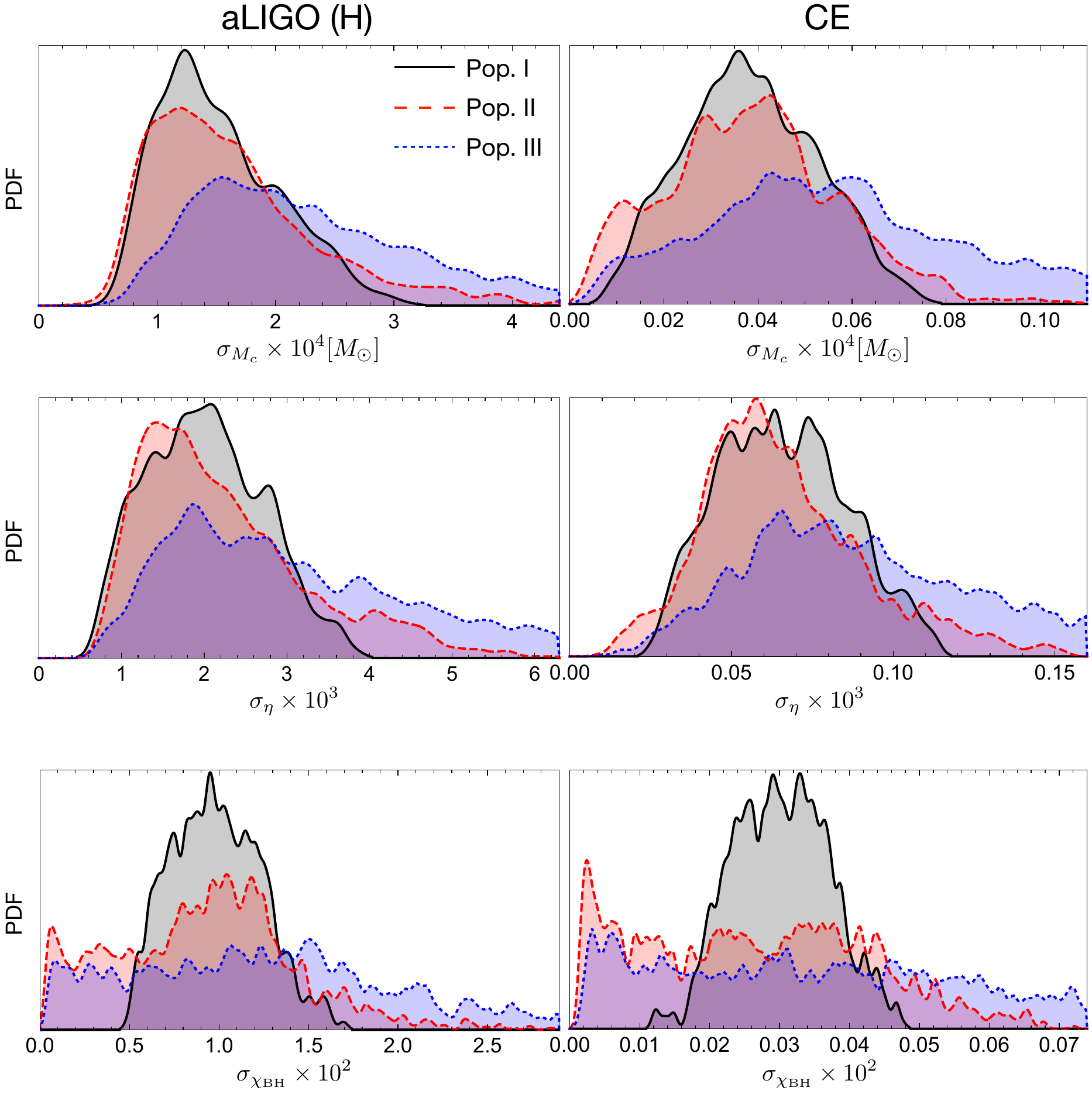}
\caption{\label{fig.pop123-errors} PDFs of $\sigma_{M_c}$ (top), $\sigma_{\eta}$ (middle), and $\sigma_{\chi_{\rm BH}}$ (bottom) for the sources in Pop-I, Pop-II, and Pop-III, obtained by the single-detector analysis.}
\end{center}
\end{figure}

\begin{figure}[t]
\begin{center}
\includegraphics[width=\columnwidth]{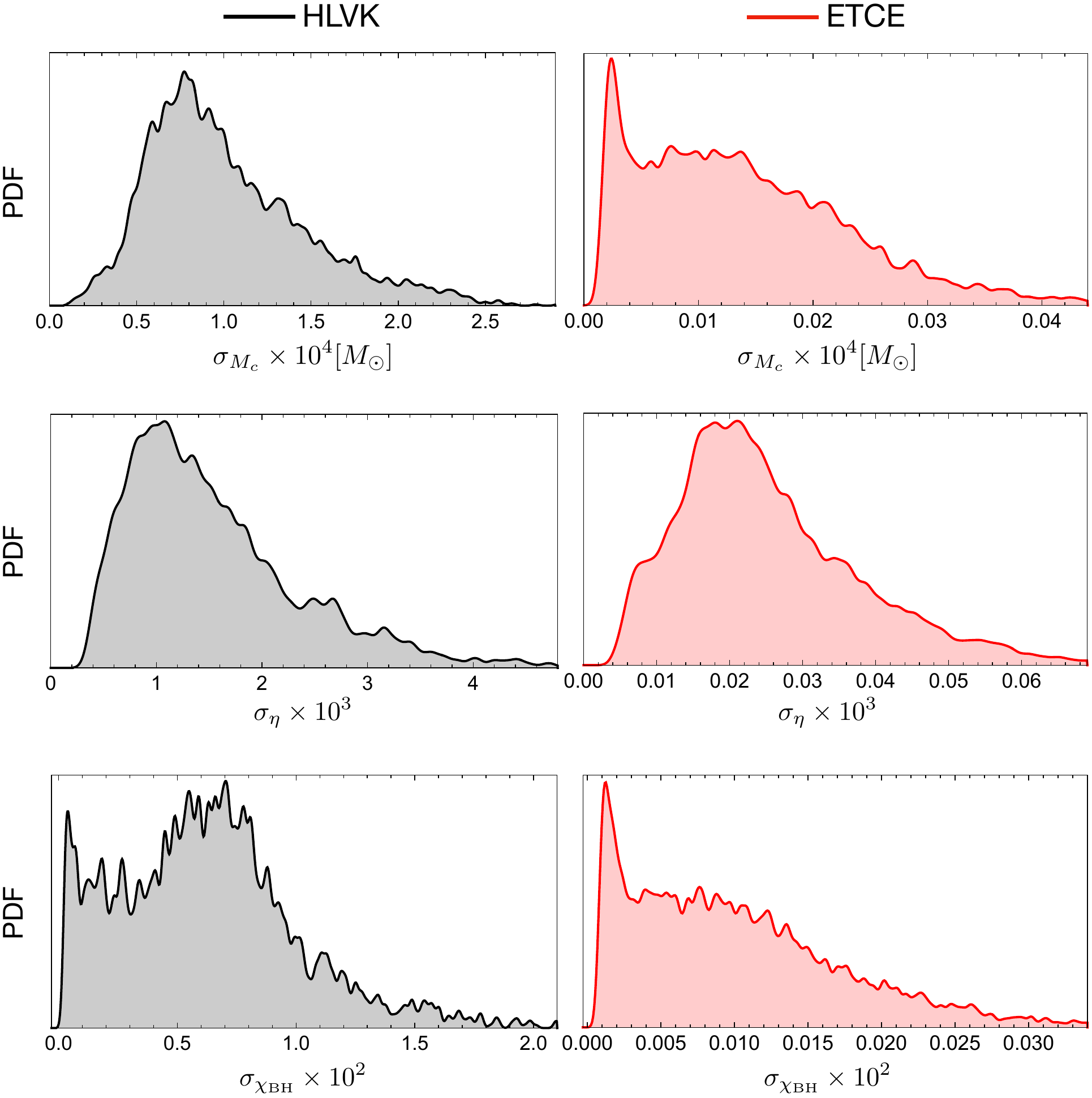}
\caption{\label{fig.Mc-eta-chi-error-histogram-3} PDFs of $\sigma_{M_c}$ (top), $\sigma_{\eta}$ (middle), and $\sigma_{\chi_{\rm BH}}$ (bottom) for the sources in Pop-III obtained by the multi-detector analysis.}
\end{center}
\end{figure}

In Fig. \ref{fig.pop123-errors}, we present the results of the chirp mass, the symmetric mass ratio, 
and the BH spin for the sources in Pop-I, Pop-II, and Pop-III,
obtained by the single-detector analysis using aLIGO (H) and CE.
In Fig. \ref{fig.Mc-eta-chi-error-histogram-3}, we also present the results of the three parameters 
for the sources in Pop-III obtained by the multi-detector analysis using the 2G and the 3G networks.

\section{ACCURACY OF THE MATRIX INVERSION} \label{ap.inversion}

We examine the inversion accuracy of our covariance matrix
by measuring the matrix norm
\be
\epsilon \equiv || \Gamma \cdot \Sigma - I  ||_{\rm max},
\ee
where $I$ is the identity matrix.
If $\epsilon$ is larger than a certain threshold, 
the covariance matrix cannot be trusted.
We empirically selected the threshold $10^4$ in this work (cf. \cite{PhysRevLett.125.201101}).
We have verified that almost all NSBH sources used in our 6--D Fisher matrices
can satisfy $\epsilon < 10^{-4}$,
and we disregarded the sources with $\epsilon$ values beyond the threshold in our PDFs.

%=======	Bibliography     ===========================	
%
%\begin{references}

\newpage

\bibliography{biblio}

%\end{references}

\end{document}